# Heavy element abundance patterns in hot DA White Dwarfs

M.A. Barstow[1], S.A. Good[1], J.B. Holberg[2], I. Hubeny[3], N.P.Bannister[1], F.C. Bruhweiler[4]
M.R. Burleigh[1] and R. Napiwotzki[5]

[1] *Department of Physics and Astronomy, University of Leicester, University Road, Leicester LE1 7RH, UK*
[2] *Lunar and Planetary Laboratory, University of Arizona, Tucson, AZ 85721, USA*
[3] *NASA Goddard Space Flight Center, Code 681, Greenbelt, Maryland, MD 20711 USA*
[4] *Institute for Astrophysics & Computational Sciences (IACS), Dept. of Physics, The Catholic Univ. of America, Washington DC 20064, USA*
[5] *Dr. Remeis-Sternwarte, Sternwartstr. 7, D-96049 Bamberg, Germany*

27 November 2002

**ABSTRACT**

We present a series of systematic abundance measurements for 25 hot DA white dwarfs in the temperature range ~20000-110000K, based on far-UV spectroscopy with STIS/GHRS on *HST*, *IUE* and *FUSE*. Using our latest heavy element blanketed non-LTE stellar atmosphere calculations we have addressed the heavy element abundance patterns making completely objective measurements of abundance values and upper limits using a $\chi^2$ fitting technique to determine the uncertainties in the abundance measurements, which can be related to the formal upper limits in those stars where particular elements are not detected.

We find that the presence or absence of heavy elements in the hot DA white dwarfs largely reflects what would be expected if radiative levitation is the supporting mechanism, although the measured abundances do not match the predicted values very well, as reported by other authors in the past. Almost all stars hotter than ~50000K contain heavy elements. For most of these the spread in element abundances is quite narrow and similar to the abundances measured in G191-B2B. However, there is an unexplained dichotomy at lower temperatures with some stars having apparently pure H envelopes and others detectable quantities of heavy elements. The heavy elements present in these cooler stars are often stratified, lying in the outermost layers of the envelope. A few strong temperature/evolutionary effects are seen in the abundance measurements. There is a decreasing Si abundance with temperature, the N abundance pattern splits into two groups at lower temperature and there is a sharp decline in Fe and Ni abundance to zero, below ~50000K. When detected, the Fe and Ni abundances maintain an approximately constant ratio, close to the cosmic value ~20. For the hottest white dwarfs observed by STIS, the strongest determinant of abundance appears to be gravity.

## 1 INTRODUCTION

The existence of two distinct groups of hot white dwarfs, having either hydrogen-rich or helium-rich atmospheres is now, qualitatively at least, understood to arise from the number of times the progenitor star ascends the red giant branch and the amount of H (and He) lost through the successive phases of mass-loss. Consequently, it seems clear that each group descends from their respective proposed progenitors, the H-rich and He-rich central stars of planetary nebulae (CSPN). Nevertheless, there remain several features of the white dwarf cooling sequence that cannot be readily explained. For example, while the hottest H-rich DA white dwarfs outnumber the He-rich DOs by a factor 7 (Fleming *et al*. 1986), the relative number of H- and He-rich CSPN is only about 4:1 (Napiwotzki 1999). In addition, there exists a gap in the He-rich track between ≈45000K and 30000K between the hot DO and cooler DB white dwarfs (Wesemael et al. 1985; Liebert et al. 1986), confirmed by a detailed spectroscopic analysis of all the then known hot He-rich objects by Dreizler and Werner (1996).

To understand white dwarf evolution, we need to know accurately several physical parameters for each star. For example, a measurement of effective temperature ($T_{eff}$) establishes how far along its evolutionary sequence the star has progressed. A key result has been the establishment of the temperature scale for DA stars through the determination of $T_{eff}$ from the Balmer line profiles in optical spectra. Bergeron Saffer & Liebert (1992) were the first to apply the technique to a large sample of white dwarfs. Combining, their results with the white dwarf evolutionary models of Wood (1992), they were able to study the mass distribution of these stars in detail. Subsequent studies have taken into account the inclusion of thin external layers of non-degenerate H and He in the evolutionary models (Wood 1995). Although the Bergeron et al. (1992) sample was constructed from a somewhat heterogeneous group of stars in the temperature range downward from ~40000K, few objects in excess of 25000K were included. Yet it is the hotter objects that are the most important test of the evolutionary models, having the greatest departure from the zero temperature Hamada & Salpeter (1961) mass-radius relation. The Palomar Green (PG) survey includes a large number of white dwarfs with effective temperatures extending up to 80000K or so, while the EUV-selected samples of Marsh et al. (1997) and Vennes et al. (1997) contained a majority of white dwarfs with $T_{eff}$ above 25000K.

The reliability of this work depends on the assumption that the Balmer line profile technique is a reliable estimator of $T_{eff}$ in all cases. Bergeron et al. (1994) demonstrated that spectroscopically undetectable traces of He could undermine the DA temperature scale at temperatures above ≈50000K. However, Napiwotzki (1997) showed this to be an artifact of assuming LTE. Also, the lack of observable photospheric He in most hot DAs (Barstow et al. 1995, 1997) could be used to argue against this being a real problem. On the other hand, Dreizler and Werner (1993) posed the question of whether or not the ubiquitous presence of elements heavier than H or He in the hottest DAs might also alter the Balmer line profiles. They concluded that the effect would only be very small, but their early non-LTE models did not include a sufficient number of level transitions to account realistically for the observed line blanketing. Using more recent model calculations, Barstow Holberg & Hubeny (1998) have shown that the presence of substantial blanketing from photospheric heavy elements does significantly alter the Balmer line profiles at a given effective temperature. Hence, the temperature scale of the hottest, most metal-rich DA white dwarfs, realised by studies using only pure H photospheric models, cannot be viewed as reliable and must be revised taking into account the photospheric composition of each star.

Until recently very few DA white dwarfs were known to have effective temperatures above 70000K. Therefore, the proposed direct evolutionary link between H-rich CSPN and white dwarfs has hardly been explored. In particular, there have been no measurements of the photospheric composition of what might be termed super-hot DAs, to distinguish them from the cooler ranges studied in detail. The *ROSAT* and *EUVE* sky surveys discovered a few examples of such very hot stars and a detailed survey of the central stars of old PN has revealed a number of others (see Napiwotzki 1999). To date, all the studies of these objects have utilized pure H models to determine $T_{eff}$ and log $g$ and have not taken into account the possible presence of heavy elements, which might be expected to lower their perceived temperature in turn. However, apart from general indications of the presence of heavy element opacity, from the EUV and soft X-ray fluxes, no direct measurements of element abundances have been made for these stars.

Consequently, we have carried out a survey of a sample of very hot DA white dwarfs with the aim of making the first measurements of their photospheric composition. Using a new grid of non-LTE, heavy element-rich model atmospheres we have analysed new *HST* STIS and *FUSE* spectra to measure the abundances of all elements detected. We have also re-determined the values of $T_{eff}$ and log $g$ for the stars in the light of these results, showing that, as might be expected, their temperatures are lower than the values inferred from earlier pure H analyses. In the wider context of white dwarf evolution, we now have the opportunity of comparing the abundances of these very hot objects with their immediate hot white dwarf descendants. To facilitate this, we have re-examined all the archive far UV spectra for the hot DA white dwarfs, from *IUE* and *HST*, obtaining revised photospheric abundance measurements based on our new model calculations. In one case, that of PG1342+444, where neither *IUE* nor *HST* data were available we used data from the *FUSE* mission to measure abundances for those species detected in this band. *FUSE* spectra are also available for several stars in the sample. However, as the coverage of our sample is not yet complete and we are concentrating on a study of those species detected in the 1100-1750Å range covered by *IUE* and *HST*, we do not include them in the current work.

## 2 OBSERVATIONS

All the new far-UV spectroscopic observations were obtained as part of a joint *HST* STIS (cycle 8) and *FUSE* (cycle 1) program, with Table 1 listing the observation dates and instruments.

### 2.1 *FUSE* spectra

An overview of the FUSE mission has been given by Moos et al. (2000) and its on-orbit performance is described by Sahnow et al. (2000). The spectrograph is described in detail by Green, Wilkinson & Friedman (1994). Further useful information is included in the *FUSE* Observer's Guide (Oegerle et al. 1998), which can be found with other technical documentation on the *FUSE* website (http://fuse.pha.jhu.edu). In this paper, only the PG1342+444 spectrum was drawn from our *FUSE* program. We have already published an initial analysis of this star and our basic reduction and analysis of the *FUSE* spectrum is described therein (Barstow et al. 2002). The resulting spectrum, combining all exposures and spectral ranges of PG1342+444 is shown in Fig. 1. The main difference between this work and the earlier paper is that we have applied a more recent version of the *FUSE* calibration and utilized the spectral analysis technique described below to measure element abundances.

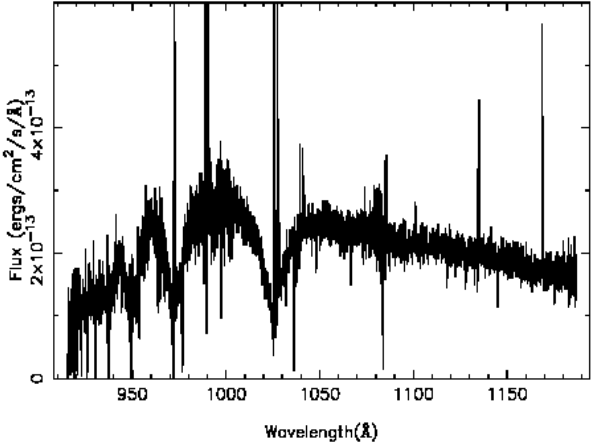

**Figure 1.** Composite *FUSE* spectrum of the hot DA white dwarf PG1342+444 produced by co-adding the individual exposures and merging the resulting spectra for the individual detector segments.

## 2.2 STIS spectra

Our GO cycle 8 STIS spectra were all obtained with the E140M grating in the ACCUM mode and cover the wavelength range from 1150 to 1700Å. All these echelle STIS spectrogram data were processed using the IDL-based CALSTIS reduction package developed and maintained by the STIS instrument development team at Goddard Space Flight Centre. A description of the algorithm can be found in Lindler (1999; http://hires/stis/docs/calstis/calstis.html).

The CALSTIS reduction offers an excellent estimation of the actual underlying background to the spectrum and the flux at each wavelength, in each spectral order. An estimate of the background requires a proper representation of different contributions to the inter-order signal. Failure to adequately estimate these can lead to both systematic relative and absolute flux errors for both the continuum and cores of spectral lines. The inter-order signal has contributions from dark rate and scattered light from adjacent orders.

The CALSTIS reduction package constructs a scattered light model, removes it from the image then re-extracts the spectrum. This process is currently accomplished in three iterations. The original net flux produced by CALSTIS is used to create an estimated source flux distribution. This is done by correcting each spectral order for echelle blaze and merging of the orders. Using this source flux distribution, one constructs a model scattered light image incorporating reference models for the echelle scattering function, telescope PSF, detector halo, grating isotropic scattering, the cross disperser scatter function and detector ghosts. The iteratively derived model scattered light image is subtracted from the original flat-fielded count rate image. The resulting net spectral image with the scattered light removed, determined for the spectrum for light incident within 5 pixels of the spectrum center, is then used to provide the resulting calibrated spectrum. A detailed discussion of this procedure is to appear elsewhere (Bowers & Lindler, in prep.). Figs. 2 and 3 show sample regions of the STIS spectrum of REJ0558-371, illustrating the detection of Nv, Ov, SiIV and FeV.

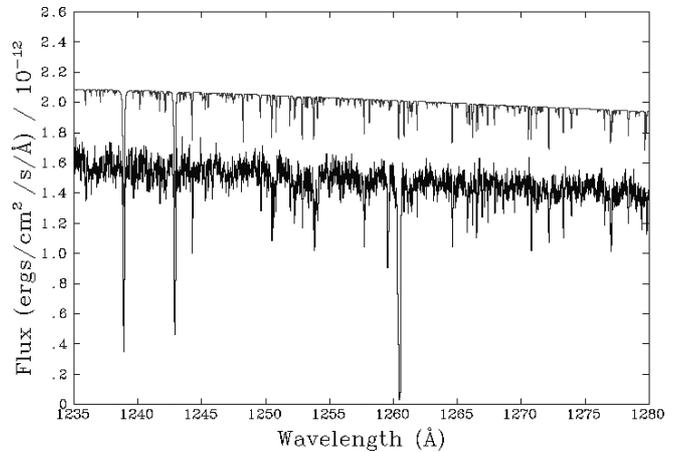

**Figure 2.** 1230Å to 1280Å region of the STIS spectrum of REJ0558-373, showing photospheric absorption lines of Nv (1238.821/1242.804Å) and large numbers of Ni lines. The best-fit synthetic spectrum is shown offset for clarity. The strong line near 1260Å, present in the observation but not in the model, is interstellar SiII.

## 2.3 Far-UV archival data

Many high dispersion far UV observations of hot DA white dwarfs have now been made with both *IUE* and *HST*. A complete examination of the reprocessed NEWSIPS white dwarf spectra has been carried out by Holberg, Barstow & Sion (1998), examining individual spectra and co-adding the multiple exposures of particular targets. However, while they discussed the detection of photospheric species, along with interstellar and circumstellar features, and their relative velocities, no abundance measurements were included. *HST* observations have been made with both the Goddard High Resolution Spectrograph (GHRS) and the Space Telescope Imaging Spectrograph (STIS). In principle, with a much larger effective area and modern electronic detectors, *HST* achieves a higher signal-to-noise than *IUE*, even when the latter spectra are co-added. On the other hand, the GHRS wavelength coverage was restricted by the spectrometer design, with several separate exposures required at different grating angles to observe all photospheric lines of interest. STIS couples high signal-to-noise with the broad wavelength coverage of the *IUE* spectra but, apart from the new observations reported in this paper, only a few exposures duplicate what is available from *IUE* or the GHRS.

**Table 1.** Summary of all far UV observations used in this paper. (H1998=Holberg et al. 1998)

| Star | Instrument | Obs. date/reference |
|---|---|---|
| PG0948+534 | STIS-E140M | 2000 April 20 |
| Ton21 | STIS-E140M | 2000 October 16 |
| REJ1738+665 | STIS-E140M | 1999 June 25 |
| PG1342+444 | *FUSE* | Barstow et al. 2002 |
| REJ2214-492 | *IUE* | Holberg et al. 1998 |
| Feige 24 | STIS-E140M | Vennes et al. 2000 |
| REJ0558-373 | STIS-E140M | 2000 April 14 |
| WD2218+706 | STIS-E140M | 1999 October 3 |
| REJ0623-371 | *IUE* | Holberg et al. 1998 |
| PG1123+189 | STIS-E140H | 1999 May 27 |
| REJ2334-471 | *IUE* | Holberg et al. 1998 |
| G191-B2B | STIS-E140M+H | Sahu et al. 1999 |
| HS1234+482 | *IUE* | Holberg et al. 1998 |
| GD246 | *IUE*/STIS-E140H/M | H1998/1998 November 20 |
| REJ0457-281 | *IUE* | Holberg et al. 1998 |
| HZ43 | *IUE*/STIS-E140H | H1998/2001April-May |
| REJ2156-546 | STIS-E140H | 1999 May 23 |
| REJ1032+532 | STIS-E140M | Holberg et al. 1999 |
| PG1057+719 | GHRS | Holberg et al. 1997 |
| GD394 | *IUE*/GHRS | Barstow et al. 1996 |
| GD153 | *IUE*/STIS-E140H | H1998/2001March |
| REJ1614-085 | GHRS | Holberg et al. 1997 |
| GD659 | *IUE*/STIS-E140H | H1998/1999 January 14 |
| EG102 | *IUE* | Holberg et al. 1998 |
| Wolf1346 | *IUE* | Holberg et al. 1998 |

We have adopted the following approach to select data from the various archival sources. If a STIS spectrum is available, we have taken this as the preferred data source (e.g. for G191-B2B, Feige 24, GD659 etc.), even if *IUE* and GHRS data have also been obtained. In some cases, only the sparse wavelength coverage of the GHRS (i.e. PG1057, REJ1614) or limited signal-to-noise of *IUE* (i.e. REJ2214, REJ0623, REJ0457, REJ2334, HZ43 etc.) is available. Most of the STIS spectra were obtained with the E140M grating which gives full wavelength coverage from 1150-1700Å, but a few targets (GD246, PG1123, HZ43 and GD153) were observed with the higher dispersion E140H grating which, in a single exposure for a particular grating setting, restricts the upper wavelength limit to ~1350Å. Therefore, we do not have access to the spectral ranges covering the O v, Si iv and C iv resonance lines in these stars. Good co-added IUE spectra do exist for GD246, HZ43 and GD153, which we use to supplement the STIS E140H data. Table 1 summarises all the spectra used for this work. Since, in several cases we are using coadded data sets published by other authors we only give the observation identification for those which are new to this work and refer back to the original references for the rest.

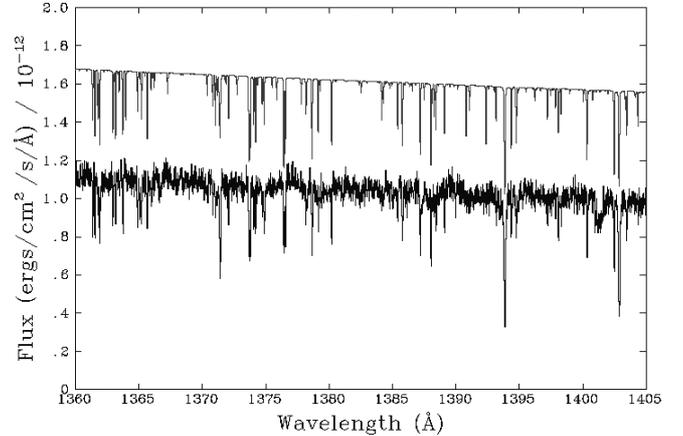

**Figure 3.** 1360Å to 1405Å region of the STIS spectrum of REJ0558-373, showing photospheric lines of O v (1371.296Å), Si iv (1393.755/1402.770Å) and Fe v. The best-fit synthetic spectrum is shown offset for clarity.

### 2.4 Optical spectra

Knowledge of the stellar effective temperature and surface gravity are important pre-requisites for determining the photospheric abundances. Values of $T_{eff}$ and log $g$ have been reported by several authors for all the stars in our sample. Yet, with improvements to model atmosphere calculations and the need to revise the measurements in the light of the stellar composition, it is necessary to repeat this work here using the existing Balmer line spectra. Most of the optical spectra we used were obtained as part of a follow-up program of spectroscopic observations to characterize the white dwarfs detected in the *ROSAT* soft X-ray and EUV sky survey.

Observations were undertaken in both Northern and Southern hemispheres. Southern hemisphere data were obtained with the 1.9-m Radcliffe reflector of the South African Astronomical Observatory (SAAO), while stars in the Northern Hemisphere were observed with the Steward Observatory 2.3-m telescope on Kitt Peak. Full details of these have already been published by Marsh et al. (1997). The main difference between the Southern and Northern Hemisphere data is their spectral resolution, 6Å (FWHM) and 8Å (FWHM) respectively. Some of our original Northern Hemisphere spectra did not cover the complete Balmer line series, excluding Hβ, due to the limited size of the CCD chip available at the time. Hence, we have supplemented the archival data with more recent observations (GD153, GD394) made using the same instrument but with a larger chip.

## 3 MODEL STELLAR ATMOSPHERE CALCULATIONS

We have calculated a new grid of model stellar atmospheres using the non-LTE code TLUSTY (Hubeny & Lanz 1995). These are based on work reported by Lanz et al. (1996) and Barstow, Hubeny & Holberg (1998, 1999). In this case we have extended the temperature range of the calculations up to 120000K, to encompass the hotter DA stars included in this analysis. To take account of the higher element ionization stages that are likely to be encountered in these objects, we have added new ions of OVI, FeVII/VIII and NiVII/VIII to the model atoms as well as extending the data for important ions such as CIV to include more energy levels. As before, all the calculations were performed in non-LTE with full line-blanketing. We initially fixed the abundances of the heavy elements at the values determined from our earlier analysis of G191-B2B, using homogeneous models (He/H = $1.0\times10^{-5}$, C/H = $4.0\times10^{-7}$, N/H = $1.6\times10^{-7}$, O/H = $9.6\times10^{-7}$, Si/H = $3.0\times10^{-7}$, Fe/H = $1.0\times10^{-5}$, Ni/H = $5.0\times10^{-7}$), but taking into account that the CIV lines near 1550Å have subsequently been resolved into multiple components by STIS (Bruhweiler et al. 2000).

In the spectrum synthesis code SYNSPEC, we have replaced the hydrogen Stark line broadening tables of Schöning & Butler (private communication) by the more extended tables of Lemke (1997). The latter allow a more accurate interpolation of the electron density for high density environments, such as the atmospheres of white dwarfs. The spectra produced by the TLUSTY/SYNSPEC codes were recently extensively tested against the results of Koester's codes (Hubeny & Koester in preparation). The differences in the predicted spectra for $T_{eff}$ =60000K and log g=8 were found to be below 0.5% in the whole UV and optical range. Furthermore, we have found that the inaccuracy in the interpolations of the Schöning and Butler tables, together with some fine details of our treatment of the level dissolution, were the primary reason for the disagreement between the spectroscopically deduced $T_{eff}$ using TLUSTY and Koester models obtained by Barstow et al. (1998).

## 4 DATA ANALYSIS

### 4.1 Measurements of $T_{eff}$ and log $g$

The technique for determining $T_{eff}$ and log $g$, by comparing the line profiles with the predictions of synthetic spectra is well established (see Holberg, Wesemael & Basile 1986; Bergeron, Saffer & Liebert 1992 and many subsequent authors). We have described our own Balmer line analysis technique in several earlier papers (e.g. Barstow et al. 1994), but we reiterate it briefly here.

The $T_{eff}$ and log $g$ measurements were performed using the program XSPEC (Shafer et al. 1991), which adopts a $\chi^2$ minimization technique to determine the model spectrum giving the best agreement with the data. The four strongest Balmer lines (β, γ, δ, ε) are simultaneously included in the fit and an independent normalisation constant applied to each, ensuring that the result is independent of the local slope of the continuum and reducing the effect of any systematic errors in the flux calibration of the spectrum.

We have established that the apparent $T_{eff}$ and log $g$ values for any white dwarf depend on knowledge of the photospheric composition of the star (Barstow et al. 1998). Hence, it is necessary to use a model grid appropriate for a particular star. It is computationally unrealistic to produce such grids for a complete range of possible compositions, but Barstow et al. (1998) demonstrated that knowledge of the exact abundance was a secondary effect. Therefore, for this analysis we computed Balmer line profiles for both pure H and G191-B2B heavy element abundances. For all stars with pure H atmospheres and those that do contain heavy elements but have temperatures below 50000K, where the influence of heavy elements on the Balmer line profiles is weaker, we used the pure H grid with the heavy element grid applied to all other stars. The results are listed in table 2 and plotted in Fig. 4.

**Table 2.** Summary of effective temperature and surface gravity for the sample of white dwarfs included in this study.

| Star | $T_{eff}$ (K) | ± 1σ | log $g$ | ± 1σ |
|---|---|---|---|---|
| PG0948+534 | 110000 | 2500 | 7.58 | 0.06 |
| Ton21 | 69711 | 530 | 7.47 | 0.05 |
| REJ1738+665 | 66760 | 1230 | 7.77 | 0.10 |
| PG1342+444 | 66750 | 2450 | 7.93 | 0.11 |
| REJ2214-492 | 61613 | 2300 | 7.29 | 0.11 |
| Feige 24 | 60487 | 1100 | 7.50 | 0.06 |
| REJ0558-373 | 59508 | 2200 | 7.70 | 0.09 |
| WD2218+706 | 58582 | 3600 | 7.05 | 0.12 |
| REJ0623-371 | 58200 | 1800 | 7.14 | 0.11 |
| PG1123+189 | 54574 | 900 | 7.48 | 0.08 |
| REJ2334-471 | 53205 | 1300 | 7.67 | 0.10 |
| G191-B2B | 52500 | 900 | 7.53 | 0.09 |
| HS1234+482 | 51950 | 800 | 7.57 | 0.07 |
| GD246 | 51308 | 850 | 7.91 | 0.07 |
| REJ0457-281 | 50960 | 1070 | 7.93 | 0.08 |
| HZ43 | 50370 | 780 | 7.85 | 0.07 |
| REJ2156-546 | 45500 | 1085 | 7.86 | 0.10 |
| REJ1032+532 | 44350 | 715 | 7.81 | 0.08 |
| PG1057+719 | 39770 | 615 | 7.90 | 0.10 |
| GD394 | 39290 | 360 | 7.89 | 0.05 |
| GD153 | 39290 | 340 | 7.77 | 0.05 |
| REJ1614-085 | 38840 | 480 | 7.92 | 0.07 |
| GD659 | 35660 | 135 | 7.93 | 0.03 |
| EG102 | 22090 | 85 | 8.05 | 0.01 |
| Wolf1346 | 19150 | 30 | 7.91 | 0.01 |

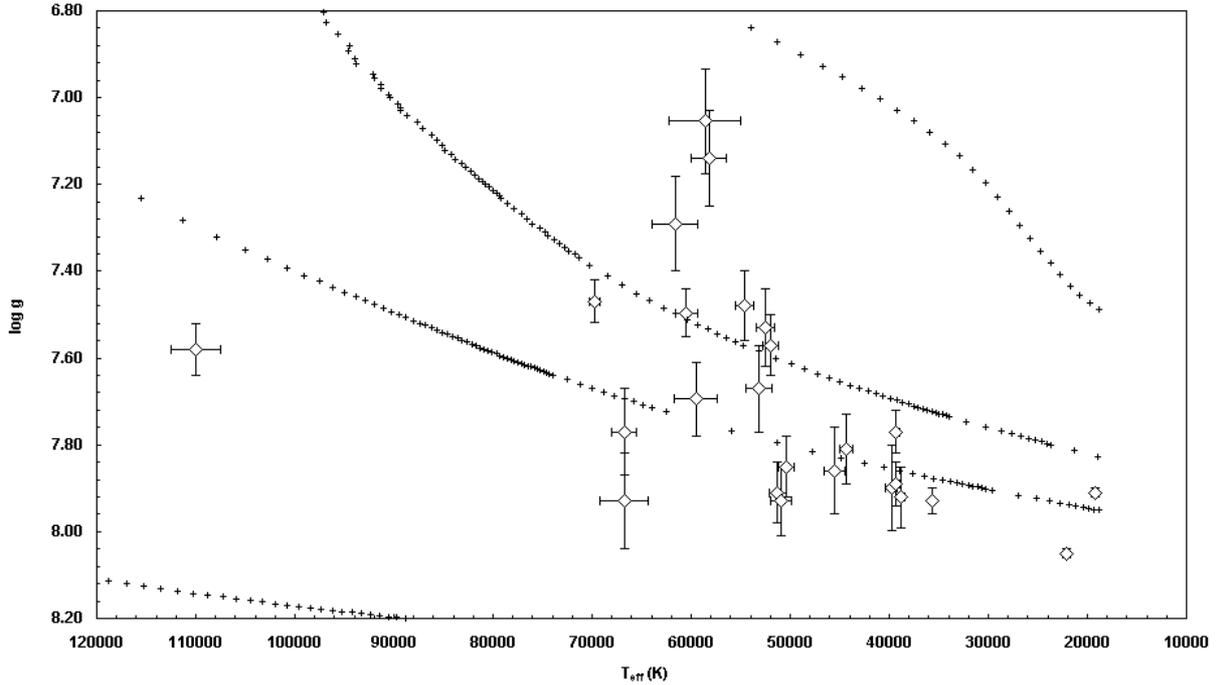

**Figure 4.** Values of $T_{\rm eff}$ and log $g$ measured from the Balmer lines for the sample of white dwarfs in this study. The crosses represent evolutionary tracks for (from top to bottom) a He core white dwarf of 0.414$M_\odot$ and C/O core masses of 0.530, 0.605 and 0.836$M_\odot$, calculated by Blöcker (1995) & Driebe et al. (1998).

### 4.2 Photospheric abundances

In far-UV studies, abundances have usually been estimated for each element in each star of the sample by visually matching the data to a synthetic spectrum calculated from the stellar model atmosphere nearest in $T_{\rm eff}$ and log $g$. Using a spectral synthesis program such as SYNSPEC, abundances can be varied within a narrow range to obtain the values that give the best representation of the data. However, it is then important to confirm the agreement by calculating a complete, fully converged, model with the adopted abundances. Although no formal goodness of fit is assigned, this approach works reasonably well when there are a small number of absorption features associated with a particular element but gets much more difficult to use when there are large numbers of lines for a particular species (e.g. FeV or NiV).

The large number of free parameters that need to be considered in studying the atmospheres of the heavy element-rich hot white dwarfs has always made a more formal, objective approach difficult to develop. Nevertheless, to understand fully the uncertainties in abundance measurements it is necessary to apply some kind of statistical analysis technique. In this paper, we fulfill this requirement by adapting the spectral fitting technique used to analyse the Balmer line spectra to examine the heavy element features in the far-UV spectra. However, a single high-resolution far-UV spectrum has a large number of data points (~60000 for the STIS echelle), which is hard for the XSPEC program to handle. Furthermore, some regions of the far-UV spectra contain little useful information, particularly at longer wavelengths. To solve both these problems we split the spectra up into smaller sections, selecting wavelength regions that contain the most useful information. These are listed in table 3, with the dominant species included.

Synthetic spectra were computed for all these wavelength ranges at each value of $T_{\rm eff}$ and log $g$ in the model grid, at the nominal abundances in the converged TLUSTY models. The spectral grid was then extended, by scaling the abundances of the main species within a particular spectral range using the SYNSPEC spectral synthesis program. Models and data were compared using the XSPEC program, considering one element at a time. When examining a spectral range containing absorption from more than a single element, the lines of any other elements were excluded from the analysis to avoid coupling the abundances together in the spectral fit. For a given star, $T_{\rm eff}$ and log $g$ were allowed to vary as free parameters, along with a single abundance scale factor, within the 1σ error ranges allowed by the Balmer analysis. In addition, 3σ uncertainties were computed for the abundance scale factors, assuming 1 degree of freedom. Although it is more usual to obtain 1σ errors, it is important to remember that the latter only correspond to a 68% probability that the true value lies within 1σ.

Furthermore, it is more common to determine 3σ limits on the abundances of non-detected elements and a comparison between such limits and a 3σ standard error is more meaningful. For reference, we determined 1σ errors for a number of the stars and note that these are typically 0.6 times the 3σ values. Examples of the fits to the Nv lines of the STIS spectrum of G191-B2B and and the noisier *IUE* spectrum of REJ0457-281 are shown in figures 5 and 6. The results of all the abundance measurements are listed in table 4.

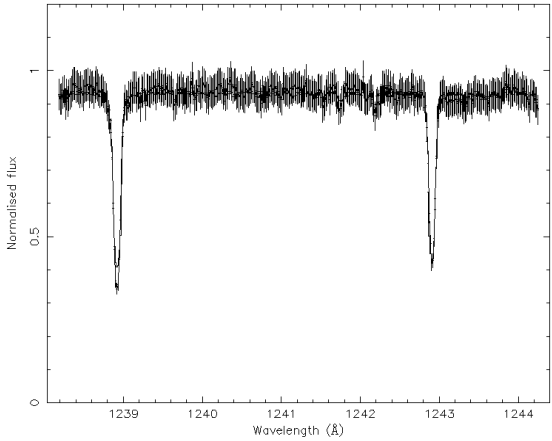

**Figure 5.** The Nv synthetic spectrum, which gives the best agreement with the STIS spectrum of G191-B2B. The data are represented by the error bars and the model by the histogram.

The absence of any entry arises mostly from a lack of data in the appropriate spectral range, rather than a true non-detection of these elements. However, in the case of PG0948+534 this stems from an inability to match the observed line profiles with the calculated models. We discuss this star in more detail later. Where a particular element was not detected, we set the abundance and lower error bound to zero, recording the 3σ upper limit in the upper error bound column.

## 5 ANALYSIS OF THE OBSERVED ABUNDANCES

The measurement of the abundances of C, N, O, Si, Fe and Ni for a sample of 25 stars, described above and listed in table 4, has generated a complex data set that requires careful study and discussion. To facilitate this process we breakdown our analysis of the results into several themes. First, we discuss each individual object in the sample in isolation, in order of decreasing $T_{eff}$, noting any peculiarities or problems with the analysis. Secondly, we consider the results for each element across the whole sample. Finally, we examine the abundances in a group of the hottest stars in our sample for which we have high signal to noise STIS spectra before making general comments about the abundance patterns we find.

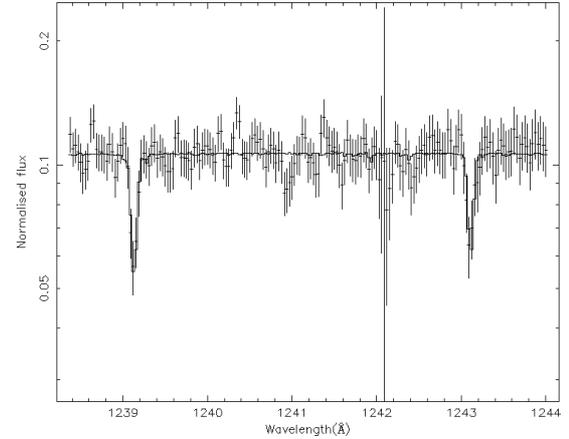

**Figure 6.** The Nv synthetic spectrum, which gives the best agreement with the *IUE* spectrum of REJ0457-281. The data are represented by the error bars and the model by the histogram.

**Table 3.** Wavelength ranges used for spectral fitting and main species included therein.

| λ (Å) | 1171 | 1235 | 1250 | 1360 | 1390 | 1450 | 1545 |
|---|---|---|---|---|---|---|---|
|  | 1181 | 1245 | 1290 | 1380 | 1405 | 1410 | 1555 |
| species | Ciii | Nv | Niv | Fev/Ov | Siiv | Fev | Civ |

### 5.1 Individual stars

#### 5.1.1 PG0948+534

This is the hottest star in the sample, by some 40000K, and the only star where the value we determine for $T_{eff}$ is not significantly reduced by the heavy element line blanketing in the models when compared with previously calculated values (Liebert, private communication). While Civ, Nv, Ov and Siiv are all clearly detected in the STIS, any attempt to fit the synthetic spectra to these features produces a poor result. The observed line strengths are much greater than the predictions of even the very highest abundances in the models (100xG191-B2B). Also, the predicted line wings are much broader than observed. A similar situation is found for the Nv lines of REJ1032+532 (see Holberg et al. 1999), which is interpreted as arising from the concentration of all the nitrogen present into the outer layers of the stellar atmosphere. The carbon, nitrogen, oxygen and silicon are probably similarly stratified in PG0948, indeed preliminary model calculations support this proposition. However, these models do not fully account for the observed features. We will consider this particular topic in a later paper. At this high temperature, Ciii is almost completely ionized and the lines of the typically strong 1170-1180Å $^3$P-$^3$P$^0$ multiplet are not detected. However, we are able to fit the Fev and Niv lines and measure the abundances of these elements.

**Table 4.** Measured abundances for each star, listed in order of decreasing $T_{eff}$, in the sample for all elements included in this study. Each group of three columns, corresponding to a particular element, lists the actual abundance value followed by the $-3\sigma$ and $+3\sigma$ uncertainties respectively. Blank cells in the table indicate that no measurements could be made, either because data were not available for the relevant spectral range or there was a technical problem with the analysis, as discussed in the text. Non-detections of a spectral feature are indicated by a value of zero in the value column and zero in the $-3\sigma$ column. The value in the $+3\sigma$ column then represents the $3\sigma$ upper limit to the abundance. All values are expressed as a number fraction with respect to hydrogen.

| Star | CIII/H | $-3\sigma$ | $+3\sigma$ | CIV/H | $-3\sigma$ | $+3\sigma$ | N/H | $-3\sigma$ | $+3\sigma$ |
|---|---|---|---|---|---|---|---|---|---|
| PG0948+534 | 0.00 | 0.00 | 4.00E-05 | | | | | | |
| Ton21 | 3.68E-07 | 3.50E-07 | 3.30E-07 | 1.71E-07 | 9.10E-08 | 1.70E-07 | 8.38E-08 | 2.30E-08 | 2.90E-08 |
| REJ1738+665 | 9.14E-08 | 8.70E-08 | 4.00E-07 | 1.34E-07 | 7.40E-08 | 1.90E-07 | 5.81E-08 | 1.70E-08 | 2.40E-08 |
| PG1342+444 | 8.80E-07 | 7.20E-07 | 3.20E-06 | 4.00E-06 | 3.60E-06 | 1.20E-05 | | | |
| REJ2214-492 | 1.00E-07 | 5.20E-08 | 1.80E-07 | 8.00E-07 | 1.70E-07 | 2.32E-07 | 1.60E-07 | 6.40E-09 | 2.72E-08 |
| Feige 24 | 7.64E-08 | 2.90E-08 | 6.00E-08 | 4.00E-07 | 4.90E-08 | 6.70E-08 | 1.60E-07 | 3.10E-08 | 2.40E-08 |
| REJ0558-373 | 1.70E-07 | 8.40E-08 | 3.40E-07 | 4.00E-07 | 3.00E-08 | 6.10E-08 | 1.86E-07 | 2.80E-08 | 5.60E-08 |
| WD2218+706 | 1.59E-07 | 1.58E-07 | 4.90E-07 | 6.70E-07 | 3.30E-07 | 2.80E-06 | 2.72E-07 | 1.20E-07 | 2.90E-07 |
| REJ0623-371 | 1.59E-07 | 1.55E-07 | 6.40E-07 | 4.00E-07 | 2.00E-07 | 4.00E-07 | 1.60E-07 | 7.00E-08 | 2.00E-07 |
| PG1123+189 | 0.00 | 0.00 | 2.80E-08 | | | | 0.00 | 0.00 | 4.00E-10 |
| REJ2334-471 | 8.00E-08 | 7.60E-08 | 1.92E-06 | 4.00E-07 | 3.00E-07 | 3.60E-07 | 2.00E-08 | 1.84E-08 | 8.00E-08 |
| G191-B2B | 1.99E-07 | 8.80E-08 | 4.40E-08 | 4.00E-07 | 9.80E-08 | 4.40E-08 | 1.60E-07 | 2.10E-08 | 4.10E-08 |
| HS1234+481 | 6.30E-08 | 5.90E-08 | 3.20E-05 | 5.00E-08 | 4.60E-08 | 4.00E-05 | 6.40E-09 | 4.80E-09 | 3.16E-07 |
| GD246 | 0.00 | 0.00 | 9.00E-09 | | | | 2.00E-09 | 4.00E-10 | 8.00E-08 |
| REJ0457-281 | 0.00 | 0.00 | 3.16E-07 | 4.00E-08 | 2.43E-08 | 1.26E-05 | 1.60E-07 | 1.35E-07 | 7.84E-06 |
| HZ43 | 0.00 | 0.00 | 2.00E-08 | 0.00 | 0.00 | 2.00E-07 | 0.00 | 0.00 | 2.00E-07 |
| REJ2156-546 | 0.00 | 0.00 | 7.30E-09 | 4.00E-09 | 2.70E-09 | 1.60E-08 | 0.00 | 0.00 | 2.50E-07 |
| REJ1032+532 | 3.00E-07 | 1.40E-07 | 5.20E-07 | 1.60E-07 | 1.20E-07 | 3.40E-07 | 5.00E-05 | 2.50E-05 | 5.00E-05 |
| PG1057+719 | | | | 0.00 | 0.00 | 1.00E-08 | 0.00 | 0.00 | 1.60E-06 |
| GD394 | 0.00 | 0.00 | 3.20E-08 | 0.00 | 0.00 | 3.20E-08 | 0.00 | 0.00 | 1.60E-06 |
| GD153 | 0.00 | 0.00 | 2.50E-08 | 0.00 | 0.00 | 4.00E-06 | 0.00 | 0.00 | 2.50E-07 |
| REJ1614-085 | | | | 4.00E-07 | 1.30E-07 | 5.00E-08 | 2.50E-04 | 1.25E-04 | 2.50E-04 |
| GD659 | 0.00 | 0.00 | 1.00E-09 | 5.00E-08 | 4.60E-08 | 2.50E-07 | 6.30E-04 | 3.20E-04 | 6.30E-04 |
| EG102 | 0.00 | 0.00 | 4.00E-06 | 0.00 | 0.00 | 4.00E-06 | 0.00 | 0.00 | 1.60E-06 |
| Wolf1346 | 0.00 | 0.00 | 5.00E-07 | 0.00 | 0.00 | 3.20E-07 | 0.00 | 0.00 | 1.60E-06 |

### 5.1.2 Ton 21

This hot DA white dwarf is included in the study of hot DA and DAO stars of Bergeron et al. (1994). In fact we use the spectrum included in that work for our determination of $T_{eff}$ and log $g$. However, nothing else other than these parameters has been published and our current work is the first spectroscopic study of its composition. All the elements considered in this study are detected and the line profiles can be matched using the synthetic spectra in the homogeneous model grid. The value of $T_{eff}$ we determine is 7500K cooler than the earlier result, which is in keeping with what we would expect in applying heavy element blanketed models rather than pure H ones (see Barstow et al. 1998). Our gravity is also about 0.3dex lower. There do not appear to be any anomalous abundances in Ton 21, compared to our reference star, G191-B2B. Therefore, we defer detailed discussion of the composition to sections 5.2, 5.3 and 5.4.

### 5.1.3 REJ1738+661

Discovered by the *ROSAT* sky survey, REJ1738+661 was the hottest white dwarf in the EUV-selected sample (Barstow et al. 1994) and may be associated with an old planetary nebula (Tweedy & Kwitter, 1994). Although the nominal value of $T_{eff}$ was reduced from ~90000K to ~70000K, when the Balmer lines were reanalysed with heavy element blanketed models (Barstow et al. 1998), it remains one of the hottest objects in our current sample. As with Ton 21, the range of models in the synthetic spectral grids can fit all the heavy element lines present and no unusual abundances are found.

| Star | O/H | -3σ | +3σ | Si/H | -3σ | +3σ |
|---|---|---|---|---|---|---|
| PG0948+534 | | | | | | |
| Ton21 | 1.21E-07 | 5.50E-08 | 8.80E-08 | 2.74E-06 | 6.30E-07 | 4.60E-07 |
| REJ1738+665 | 4.84E-07 | 1.90E-07 | 2.90E-07 | 2.15E-06 | 6.70E-07 | 2.30E-06 |
| PG1342+444 | 7.90E-06 | 5.70E-06 | 1.40E-05 | 7.50E-07 | 3.70E-07 | 7.00E-07 |
| REJ2214-492 | 2.40E-06 | 7.70E-07 | 1.00E-06 | 1.20E-06 | 1.56E-07 | 4.08E-07 |
| Feige 24 | 2.02E-07 | 2.01E-07 | 1.40E-07 | 2.31E-06 | 7.70E-07 | 2.00E-06 |
| REJ0558-373 | 2.89E-06 | 8.50E-07 | 1.80E-06 | 3.00E-06 | 1.40E-06 | 1.20E-06 |
| WD2218+706 | 9.60E-06 | 1.90E-06 | 6.20E-06 | 1.93E-06 | 1.10E-06 | 1.10E-05 |
| REJ0623-371 | 2.00E-06 | 1.80E-06 | 8.00E-06 | 7.50E-07 | 6.00E-07 | 1.30E-06 |
| PG1123+189 | | | | | | |
| REJ2334-471 | 7.10E-07 | 7.00E-07 | 1.00E-04 | 4.80E-07 | 4.00E-07 | 1.90E-06 |
| G191-B2B | 3.51E-07 | 2.00E-07 | 7.40E-07 | 8.65E-07 | 3.50E-07 | 3.20E-07 |
| HS1234+481 | 5.00E-06 | 4.99E-06 | 9.50E-05 | 1.50E-08 | 1.20E-08 | 5.85E-07 |
| GD246 | 1.60E-07 | 1.50E-07 | 1.14E-06 | 1.20E-07 | 3.50E-08 | 5.00E-08 |
| REJ0457-281 | 1.00E-05 | 9.99E-06 | 9.00E-05 | 3.00E-08 | 2.70E-08 | 2.10E-07 |
| HZ43 | 0.00 | 0.00 | 2.40E-06 | 0.00 | 0.00 | 1.20E-09 |
| REJ2156-546 | 0.00 | 0.00 | 2.40E-06 | 7.50E-10 | 3.70E-10 | 1.10E-09 |
| REJ1032+532 | 1.20E-07 | 1.19E-07 | 1.40E-06 | 9.50E-08 | 4.70E-08 | 9.50E-08 |
| PG1057+719 | | | | 0.00 | 0.00 | 4.80E-10 |
| GD394 | 0.00 | 0.00 | 4.80E-06 | 2.80E-06 | 5.00E-07 | 2.00E-07 |
| GD153 | 0.00 | 0.00 | 9.60E-06 | 0.00 | 0.00 | 2.40E-09 |
| REJ1614-085 | | | | 9.50E-09 | 8.30E-09 | 7.00E-09 |
| GD659 | 0.00 | 0.00 | 3.80E-06 | 4.80E-09 | 3.80E-09 | 7.00E-08 |
| EG102 | 0.00 | 0.00 | 2.40E-06 | 2.40E-07 | 2.36E-07 | 1.60E-06 |
| Wolf1346 | 0.00 | 0.00 | 7.60E-07 | 3.80E-09 | 3.50E-09 | 4.40E-08 |

| Star | Fe/H | -3σ | +3σ | Ni/H | -3σ | +3σ |
|---|---|---|---|---|---|---|
| PG0948+534 | 1.90E-06 | 5.00E-07 | 6.00E-07 | 1.20E-07 | 3.00E-08 | 3.00E-08 |
| Ton21 | 1.27E-06 | 4.30E-07 | 4.20E-07 | 6.64E-08 | 2.50E-08 | 5.60E-08 |
| REJ1738+665 | 2.10E-07 | 1.10E-07 | 4.60E-07 | 5.24E-08 | 4.70E-08 | 1.40E-07 |
| PG1342+444 | 1.00E-04 | 9.99E-05 | 9.00E-04 | 5.00E-06 | 4.40E-07 | 1.10E-05 |
| REJ2214-492 | 1.45E-05 | 1.82E-06 | 1.56E-06 | 1.00E-06 | 7.00E-08 | 1.20E-07 |
| Feige 24 | 3.56E-06 | 6.50E-07 | 1.40E-06 | 1.16E-07 | 3.00E-08 | 5.40E-08 |
| REJ0558-373 | 9.93E-06 | 4.00E-06 | 2.20E-06 | 6.71E-07 | 1.00E-07 | 2.10E-07 |
| WD2218+706 | 1.00E-05 | 5.00E-06 | 1.50E-05 | 9.91E-07 | 3.00E-07 | 3.80E-07 |
| REJ0623-371 | 1.58E-05 | 3.20E-06 | 1.20E-05 | 1.30E-06 | 6.00E-07 | 5.00E-07 |
| PG1123+189 | | | | 1.30E-07 | 4.00E-08 | 5.00E-08 |
| REJ2334-471 | 2.50E-06 | 2.10E-07 | 1.05E-05 | 2.00E-08 | 1.50E-08 | 9.80E-07 |
| G191-B2B | 3.30E-06 | 1.20E-06 | 3.10E-06 | 2.40E-07 | 2.40E-08 | 8.40E-08 |
| HS1234+481 | 2.00E-06 | 1.90E-06 | 1.40E-05 | 4.00E-07 | 3.96E-07 | 2.80E-06 |
| GD246 | 0.00 | 0.00 | 1.20E-07 | 7.90E-08 | 7.40E-08 | 1.30E-07 |
| REJ0457-281 | 3.20E-06 | 2.40E-06 | 1.28E-05 | 1.00E-06 | 9.00E-07 | 2.50E-06 |
| HZ43 | 0.00 | 0.00 | 1.60E-06 | 0.00 | 0.00 | 5.00E-06 |
| REJ2156-546 | 0.00 | 0.00 | 2.80E-06 | 0.00 | 0.00 | 1.50E-07 |
| REJ1032+532 | 0.00 | 0.00 | 5.00E-06 | 0.00 | 0.00 | 6.30E-07 |
| PG1057+719 | 0.00 | 0.00 | 2.00E-06 | 0.00 | 0.00 | 1.30E-06 |
| GD394 | 0.00 | 0.00 | 8.00E-08 | 0.00 | 0.00 | 1.00E-06 |
| GD153 | 0.00 | 0.00 | 1.60E-05 | 0.00 | 0.00 | 6.30E-07 |
| REJ1614-085 | 0.00 | 0.00 | 7.90E-06 | 0.00 | 0.00 | 1.30E-06 |
| GD659 | 0.00 | 0.00 | 2.00E-05 | 0.00 | 0.00 | 6.30E-08 |
| EG102 | 0.00 | 0.00 | 2.00E-05 | 1.00E-14 | 0.00 | 1.30E-07 |
| Wolf1346 | 0.00 | 0.00 | 6.30E-06 | 0.00 | 0.00 | 1.60E-08 |

*5.1.4 PG1342+444*

PG1342+444 is the only star in our sample for which neither *IUE* nor *HST* spectra are available. However, we do have a *FUSE* spectrum of the object, which has been the subject of an earlier analysis (Barstow et al. 2002). For the present work we adopted the values of $T_{\mathrm{eff}}$ and log $g$ determined from the Balmer rather than the Lyman lines, which give an unexpectedly lower value (Barstow et al. 2001c, 2002). In the original discussion, we used the argument that good agreement between the C abundances obtained from CIII and CIV lines respectively favoured the higher temperature. However, with the more rigorous statistical treatment applied here, it can be seen that the uncertainties in the abundance measurements are large and of a similar magnitude to the reported abundance discrepancies for the Lyman-determined temperature. In general, the earlier result of photospheric abundances up to an order of magnitude greater than those found for G191-B2B holds, despite the large measurement uncertainties.

### 5.1.5 REJ2214-492

This is the brightest of several hot, heavy element-rich white dwarfs discovered by the *ROSAT* WFC, which appeared to be very similar to G191-B2B (Holberg et al. 1993, 1994). Initial abundances determined from the coadded *IUE* spectra appeared to be slightly higher than for G191-B2B, but these results depend on the assumed temperature, which has subsequently been revised downwards using improved optical spectra and heavy element blanketed models (Barstow et al. 1998). In the present study C, O, Fe and Ni seem to be enhanced when compared to G191-B2B.

### 5.1.6 Feige 24

Feige 24 is a well-studied hot, heavy element rich white dwarf, very similar to G191-B2B and the STIS observation of has been reported already by Vennes et al. (2000) and Vennes & Lanz (2001). Feige 24 is the only known close binary in the sample. The analysis of Vennes & Lanz reports that the average heavy element abundance in Feige 24 is 0.17 dex larger than for G191-B2B, apart from for carbon, which is more abundant in the latter star. However, we find little difference between the abundances of Feige 24 and G191-B2B, within the formal errors of our analysis (Fig. 7). Indeed, the abundances measured from CIV, NV and FeV are in very close agreement. Only the Si and Ni abundances disagree by amounts greater than the 3σ uncertainties. We note that the measured abundances are sensitive to the adopted temperature. Our value of 60500K is ~3000K hotter than that of Vennes & Lanz, which may explain the difference between these results.

### 5.1.7 REJ0558-367

The current paper represents the first detailed analysis of the element abundances, apart from the preliminary work reported by Barstow et al. (2001). It has a similar composition to G191-B2B.

### 5.1.8 WD2218+706

This object has also been discussed in some detail individually (Barstow et al. 2001a) and is known to have a planetary nebula. Unlike any of the other white dwarfs in the sample, the STIS spectrum reveals a trace of HeII at 1640Å. Hence, this white dwarf is a DAO, although the He abundance is too low to be detected at 4686Å, in the visible band. This may be related to the fact that this star has the lowest gravity (7.05) and mass (0.45-0.5M$_\odot$) of any in the current sample. There is currently no evidence for a binary companion. The first analysis indicated that N and O were enhanced compared to G191-B2B, with the other element abundances lying within a factor 2. However, only O seems to be significantly more abundant in the current work.

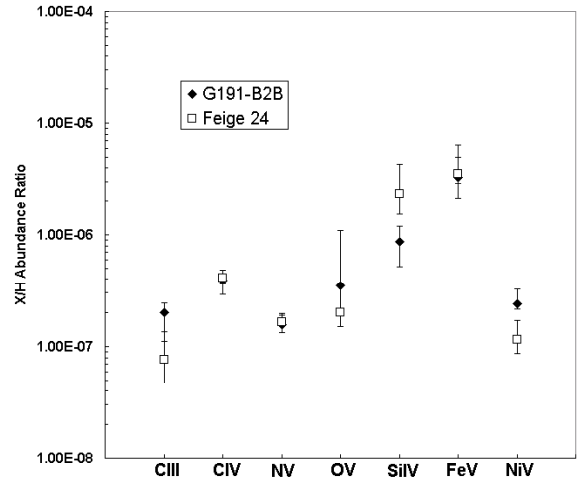

**Figure 7.** Comparison of measured abundances for Feige 24 and G191-B2B.

### 5.1.9 REJ0623-371

REJ0623-371 is another G191-B2B analogue discovered by ROSAT (Holberg et al. 1993). The abundances measured here seem to be similar to those of G191-B2B.

### 5.1.10 PG1123+189

This white dwarf is a member of a group of objects that contain lower heavy element abundances than the G191-B2B-like group, on the basis of its EUV flux (see Wolff et al. 1998). However, the E140H echelle spectrum included in this work has limited spectral coverage (1170-1360Å), including only NV and NiIV lines. Nitrogen is not detected, but Ni appears to be present. Optical spectroscopy by Schmidt & Smith (1995) reveals a cool companion longward of 6000Å. Radial velocity measurements (Saffer, Livio & Yungelson 1998) indicate that the pair are not close.

### 5.1.11 REJ2334-471

Another ROSAT heavy element-rich discovery, fainter than both REJ2214-492 and REJ0623-371, the coadded *IUE* spectrum has lower signal-to-noise than these visually brighter stars. Surprisingly, no detailed analysis of this data has been published before, although the detection of CIV, NV and SiIV are noted by Holberg et al. 1998. We confirm the presence of these features and obtain limits on the abundance of O, Fe and Ni. Interestingly, the best-fit model yields a finite abundance of Fe at a level ~10 times the −3σ error bar. As no individual Fe lines can be positively detected, it seems likely that their strength is very similar to the noise level.

*5.1.12 G191-B2B*
G191-B2B is the visually brightest and the best studied of all hot white dwarfs. It has been observed in many wavebands, providing a calibration standard and reference against which we compare all related objects, as can be seen in the above sections. Far-UV abundance measurements have been reported by many authors, using *IUE*, FOS, GHRS and STIS spectra (e.g. Vennes et al. 1992; Sion et al. 1992; Vidal-Madjar et al. 1994; Vennes & Lanz 2001). Nevertheless, the star continues to spring a few surprises. Bruhweiler et al. (2000) and Vennes & Lanz note that the CIV resonance doublet near 1550Å has an additional component that is blue-shifted with respect to the photospheric velocity. The nature of the component is not important for this discussion and will be reported elsewhere (Bannister et al. 2003). However, it has led to erroneously high estimates of the C abundances based on *IUE* spectra, where the components cannot be resolved. Consequently, all C abundances based on CIV lines observed by *IUE* should be treated with caution.

The importance of this new measurement of the photospheric abundances for G191-B2B is the application of the objective technique outlined in this paper and the formal errors that can be derived. Comparing these results with the recent analysis of Vennes & Lanz (2001) shows a reasonable agreement. The values we obtain are not precisely the same, but are more or less consistent within the overall uncertainties imposed by the signal-to-noise of the spectra and the possible range of adopted temperatures.

*5.1.13 HS1234+481*
Little is known about the photospheric abundance pattern in this star. Jordan et al. (1996) have reported the detection of Fe in the *EUVE* spectrum. Unfortunately, the single *IUE* echelle spectrum is too noisy to reveal very much and no heavy element detections can be reported. The upper limits are included in table 4 for completeness, but the error bars too large to provide any useful information. Consequently, this star is not included in any further analysis or discussion of the sample.

*5.1.14 GD246*
Vennes, Thejll & Shipman (1991) reported the presence of the CIV, NV and SiIV resonance lines and derived abundances for these elements. However, the coadded *IUE* spectrum does not reveal the N features (Holberg et al. 1998). They are weak, but detected in the STIS E140H echelle data. We are only able to place upper limits on the presence of O, Fe and Ni. However, Vennes & Dupuis (2002) report the detection of FeVI features in the Chandra spectrum of this star.

While we can obtain abundances for both N and Si, our analysis of the CIV doublet poses a problem. It is not possible to match the strength and shape of the observed features with the model spectra. Given, the good agreement achieved for most of the stars we have studied, this is a surprise and reminiscent of the problem encountered for all the resonance lines detected in PG0948+534 and discussed above. This suggests that the CIV lines contain a second absorbing component, in addition to the photospheric material, that is not resolved by the *IUE* or more recent STIS E140M spectra. This problem would not have been noted by Vennes et al. (1991), as their abundance analysis was based on measurements of equivalent width rather than line profile fitting.

*5.1.15 REJ0457-281*
Multi-wavelength observations of the star indicate that is it very similar to G191-B2B. However, the co-added *IUE* spectrum reveals only C, N and Si lines (see Holberg et al. 1998) and O, Fe and Ni have not been detected. However, the upper limits we obtain are consistent with the star having the same abundances of these elements as G191-B2B, the noise level in the spectrum being similar to the predicted line strengths. Bannister et al. (2003) note that the CIV doublet has a blue-shifted component, as in G191-B2B, but the velocity separation is larger and the lines are well resolved.

*5.1.16 HZ43*
HZ43 is well known as a pure H atmosphere white dwarf. It is relatively hot at 50400K, only ~2000K cooler than G191-B2B. Hence, it is a surprise that no sources of photospheric opacity have been detected in the soft X-ray, EUV, far UV and visible observations and only upper limits to abundances obtained (e.g. Napiwotzki et al. 1993; Barstow, Holberg & Koester 1995). Our analysis includes both the earlier *IUE* data and a recent high signal-to-noise STIS E140H spectrum recorded for the purpose of studying the interstellar deuterium and covering ~1170-1360Å. No photospheric heavy elements are detected in the STIS spectrum. However, we provide an objective determination of the upper limits to the abundances for the first time. We note that Dupuis et al. (1998) report upper limits on C, N and O from *ORFEUS* data, which are similar to those derived here.

*5.1.17 REJ2156-546*
EUV studies of this star indicate that it has a pure H atmosphere (Barstow et al. 1997). Its primary interest for far-UV observations has been as a background source for exploring the interstellar medium, particularly as the EUV spectrum indicates that the line of sight is significantly ionized (Barstow et al. 1997). The STIS E140M spectrum largely confirms absence of photospheric material, although there appears to be a weak detection of the shorter wavelength component of the CIV doublet.

*5.1.18 REJ1032+532*
This object is an interesting contrast to REJ2156-546. Although their temperatures and gravities are almost identical, the STIS E140M spectrum of REJ1032+532 reveals photospheric lines of CIII, CIV, NV, SiIII and SiIV (Holberg et al. 1999). However, the EUV spectrum indicates that the atmosphere is pure hydrogen (Barstow et al. 1997). As discussed by Holberg et al. (1999), this discrepancy can be explained if the nitrogen is not homogeneously mixed in the atmosphere but stratified, residing mainly in the outermost regions of the atmosphere. The nitrogen abundance in table 4 is, therefore, derived from the stratified model of Holberg et al. We also note that, although we formally get a good fit to the CIV lines, the models cannot quite reproduce the cores, indicating that the carbon may also be stratified to a degree.

*5.1.19 PG1057+719*
PG1057+719 was observed by the GHRS, with several central wavelength settings yielding coverage of CIV, NV, SiIV, Fe and Ni, but not CIII or OV (Holberg et al. 1997). No heavy elements were detected in this star and our analysis provides just upper limits.

*5.1.20 GD394*
The nature of the far-UV opacity observed in GD394 was the subject of some uncertainty for a while. However, this issue was resolved by Barstow et al. (1996), concluding that the observed Si is photospheric. No other elements are detected in the *IUE* or GHRS spectra, although the EUV spectrum indicates that there is a greater photospheric opacity than can be accounted for by Si alone. Subsequently, Chayer et al. (2000) have detected FeIII in *FUSE* data, while Dupuis et al. (2000) found the white dwarf to be photometrically variable in the EUV and observed an extreme abundance of Si in the optical. Episodic accretion is proposed as the origin of a large EUV-dark spot on the surface, which may also explain the anomalous Si abundance. In any interpretation of the existing far UV data, the possibility that the features are time averaged over the 1.150 day EUV cycle or may be variable in other ways must be taken into account. At the moment the sources of the accreted material remains unknown.

*5.1.21 GD153*
A pure H atmosphere white dwarf that is often used as a calibration source, GD153 was observed by the STIS 140H echelle for the purpose of studying interstellar deuterium. No photospheric heavy elements are detected in this star.

*5.1.22 REJ1614-085*
A close twin to PG1057+719, REJ1614-085 was observed as part of the same GHRS program (Holberg et al. 1997). In contrast to PG1057+710, photospheric heavy elements were detected in this star. Our current analysis provides revised abundance measurements for C, N and Si, together with upper limits for Fe and Ni. Infra-red spectroscopy from 1-2.5μm by Burleigh et al. (2003) rules out the presence of any cool companion earlier than M7V.

*5.1.23 GD659*
Heavy elements detected in the *IUE* spectrum were originally attributed to circumstellar material (Holberg et al. 1995). A more recent measurement of the stellar radial velocity indicates that the observed features are in fact photospheric. This poses an interesting problem, since the EUV spectrum shows a pure H atmosphere (Barstow et al. 1997). Furthermore, a good fit to the NV lines cannot be achieved with our models. It appears that GD659 has a stratified N component, like REJ1032+532. Therefore, the N abundance listed in table 4 was determined under this assumption (as for REJ1032+532). We note that the "circumstellar" C, N and Si lines listed for this star in Holberg et al. (1998) are incorrectly attributed and should be included in the "photospheric" list.

*5.1.24 EG102*
This star is often used as a flux standard. Vennes et al. (1991) reported the first detection of SiIII in this star. Holberg et al. (1998) note that photospheric Al is also present. A strong CII line at 1335Å is considered to be circumstellar. Magnesium has also been detected at visible wavelengths. The short downward diffusion time for Mg and Al would indicate that the object is undergoing accretion from an unknown source. In this paper, we have not generally included Mg and Al in the models and only report our measurement of the Si abundance.

*5.1.25 Wolf 1346*
At 19200K, Wolf 1346 is the coolest star in our sample. Detection of SiII was first reported by Bruhweiler & Kondo (1983) and the photospheric origin of the lines confirmed by Holberg et al. (1996).

**5.2 Observed patterns of element abundances**

The spectral analysis presented in this paper has obtained measurements of DA white dwarf atmosphere heavy element abundances in a uniform way. While the spectra do not necessarily have uniform signal-to-noise, the error bars on the measurements have been determined with a full statistical treatment and also take account of uncertainties in adopted parameters such as $T_{eff}$ and log $g$. Therefore, this represents the most comprehensive and self-consistent dataset available for studying how the atmospheric compositions of hot DA white dwarfs evolve, varying with age and mass. For the purposes of this discussion we use the parameters we can measure directly, $T_{eff}$ and log $g$, as proxies for these in Figs. 8 to 17, where we show the measured abundances and upper limits for each element in turn. Generally, a much stronger relationship between abundance and temperature is seen than for gravity. Hence, most of the following discussion will concentrate on Figs. 8, 10, 12, 14 and 16, considering each element in turn.

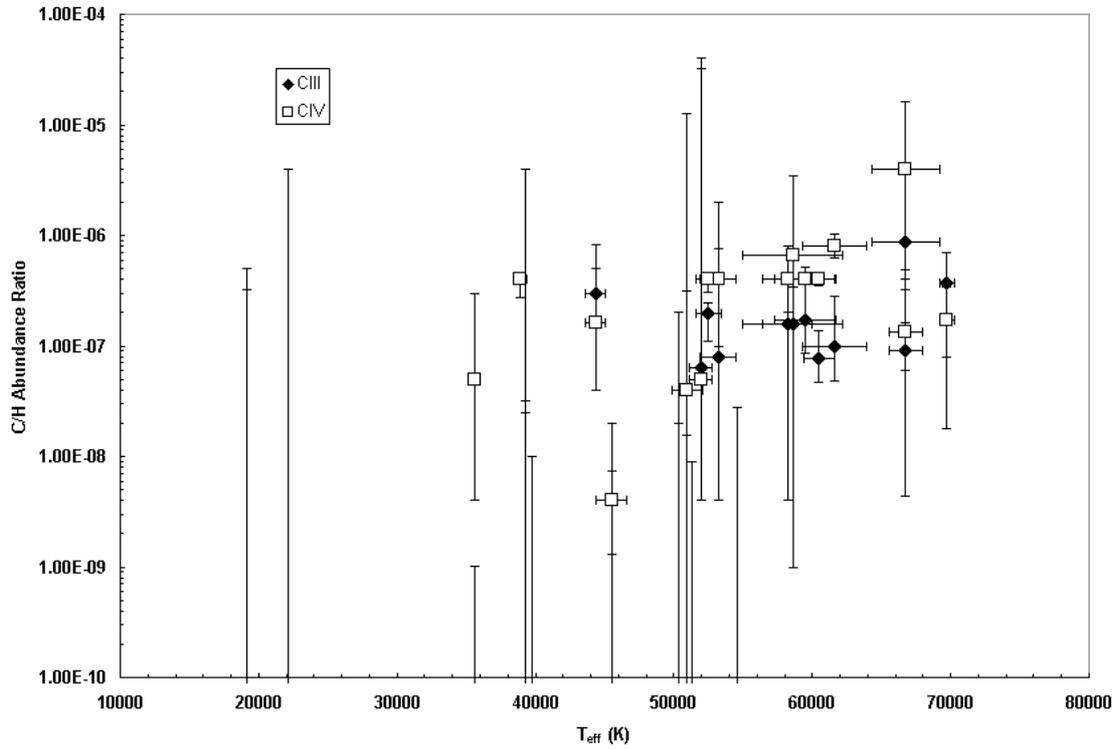

**Figure 8.** Measured abundances of Carbon (by number with respect to hydrogen) as a function of $T_{\text{eff}}$ for both CIII (filled diamonds) and CIV (open squares) lines.

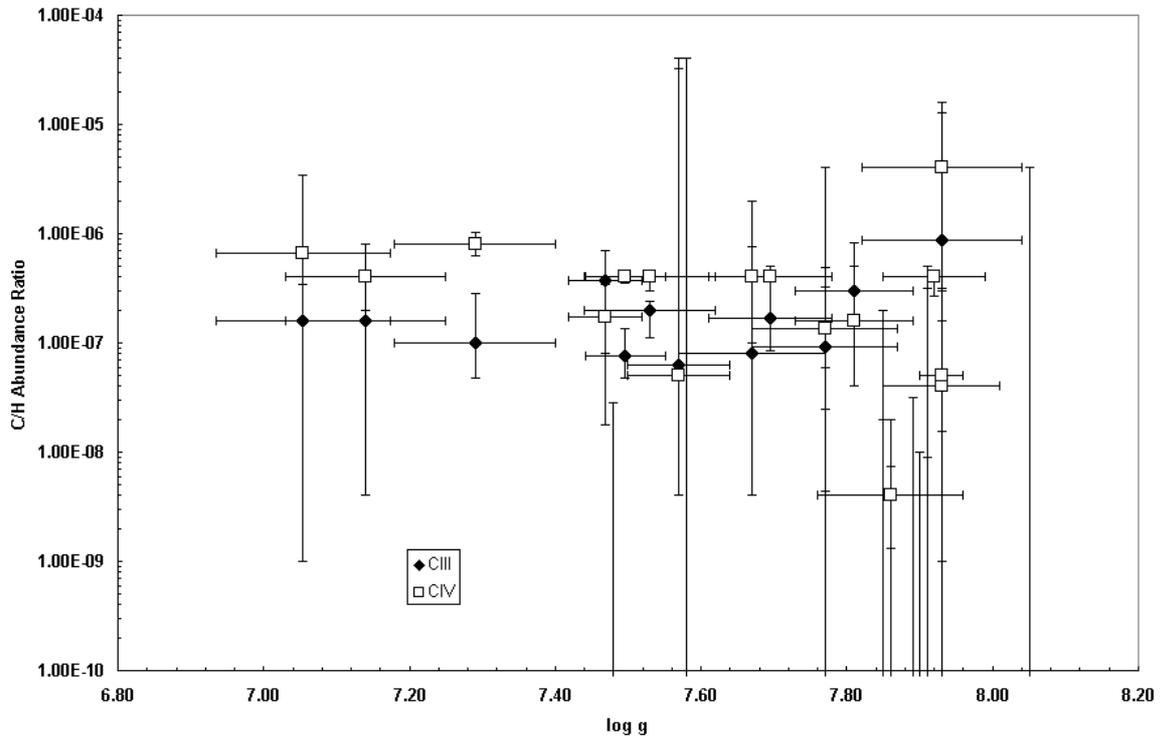

**Figure 9.** Measured abundances of Carbon (by number with respect to hydrogen) as a function of $\log g$ for both CIII (filled diamonds) and CIV (open squares) lines.

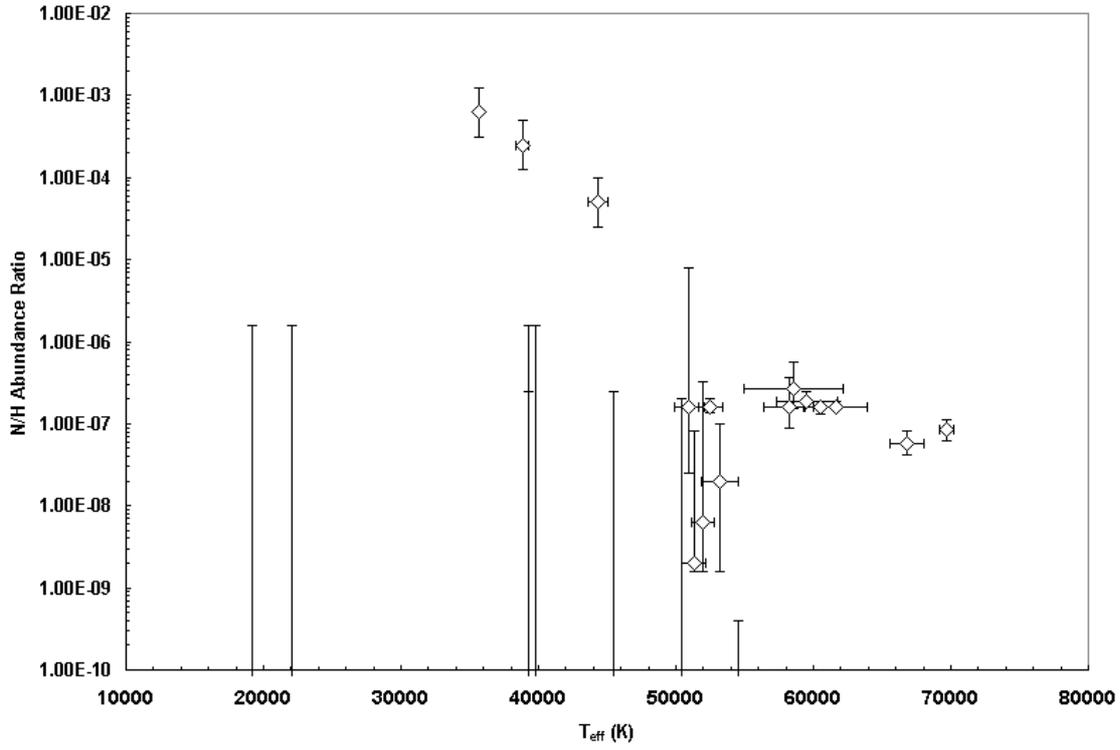

**Figure 10.** Measured abundances of Nitrogen (by number with respect to hydrogen) as a function of $T_{\text{eff}}$.

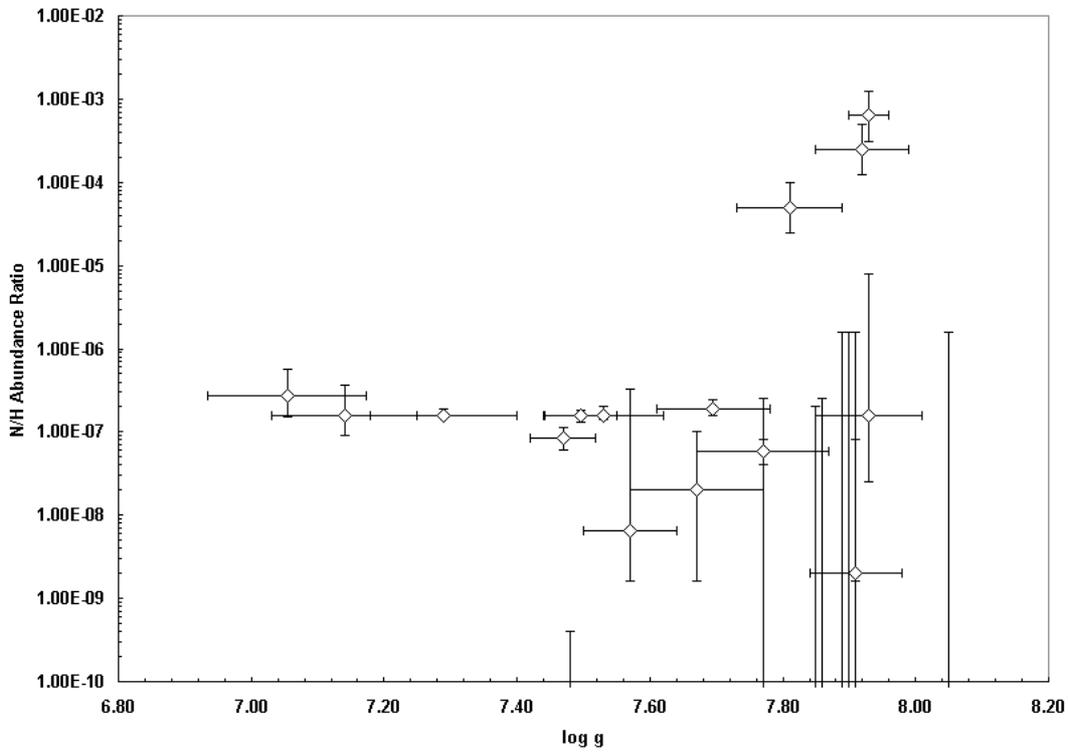

**Figure 11.** Measured abundances of Nitrogen (by number with respect to hydrogen) as a function of log $g$.

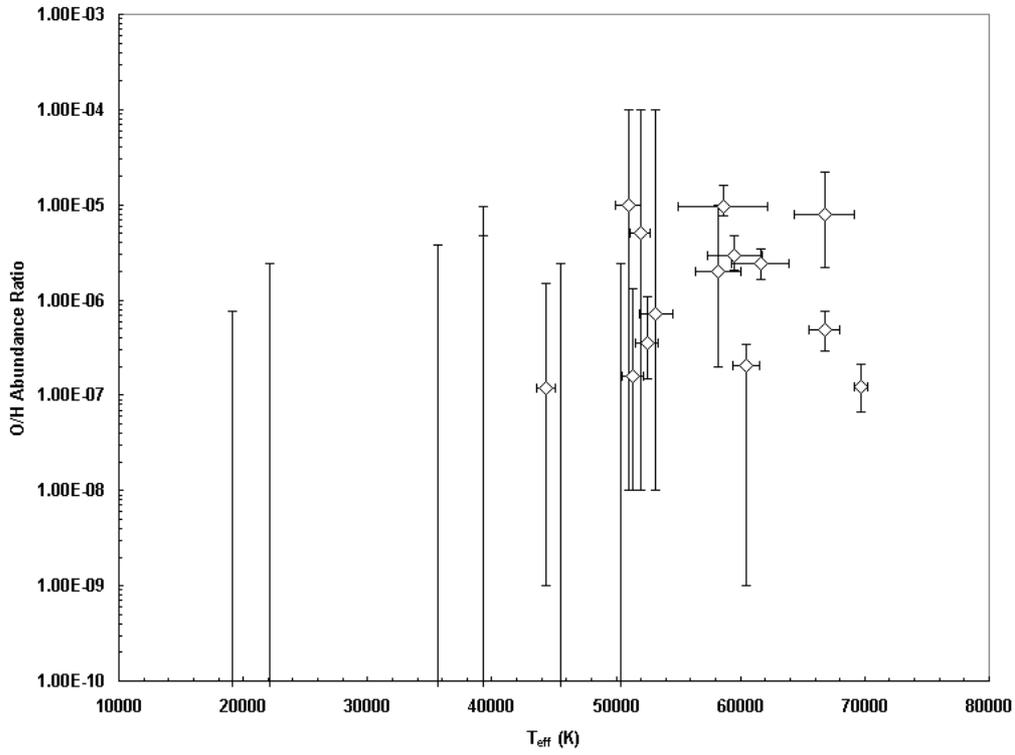

**Figure 12.** Measured abundances of Oxygen (by number with respect to hydrogen) as a function of $T_{\text{eff}}$.

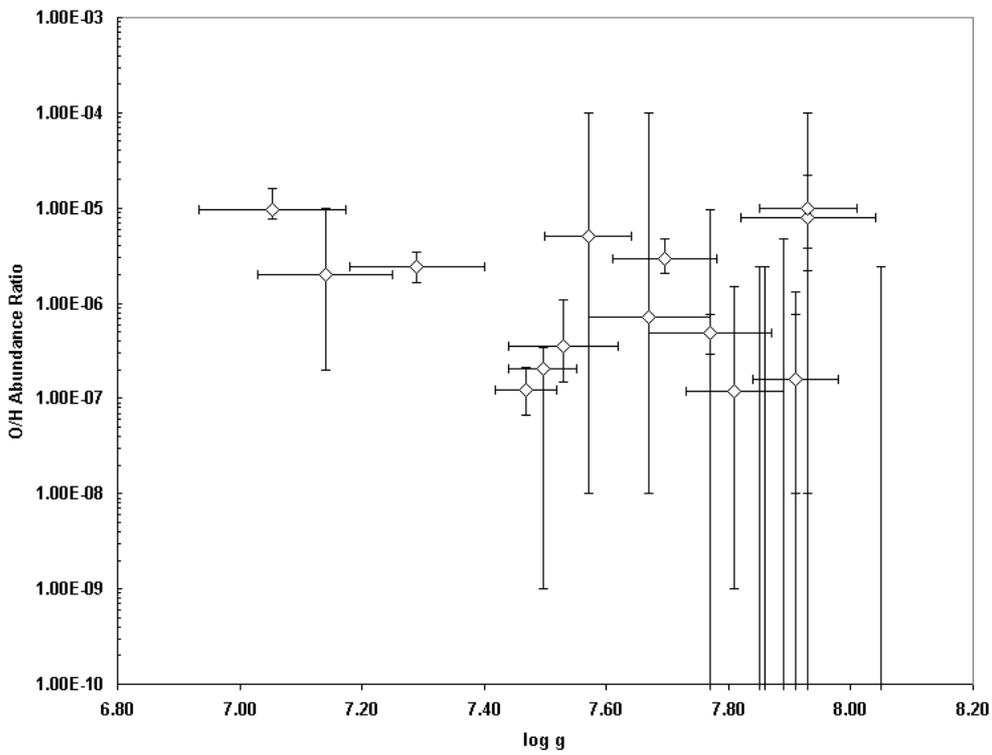

**Figure 13.** Measured abundances of Oxygen (by number with respect to hydrogen) as a function of log $g$.

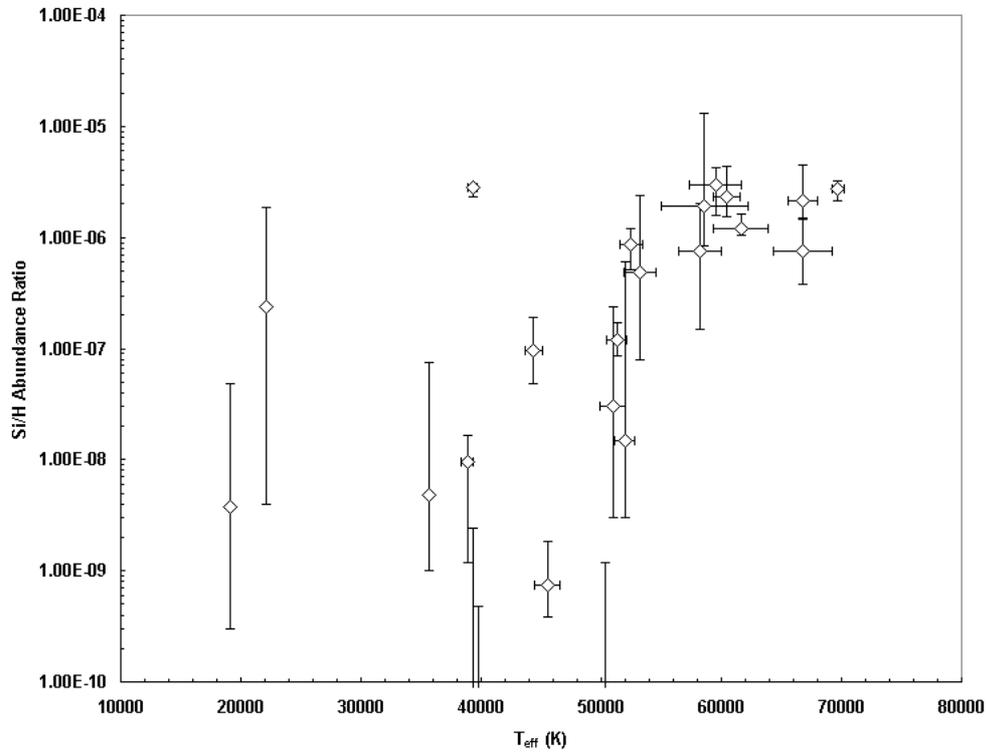

**Figure 14.** Measured abundances of Silicon (by number with respect to hydrogen) as a function of $T_{\rm eff}$.

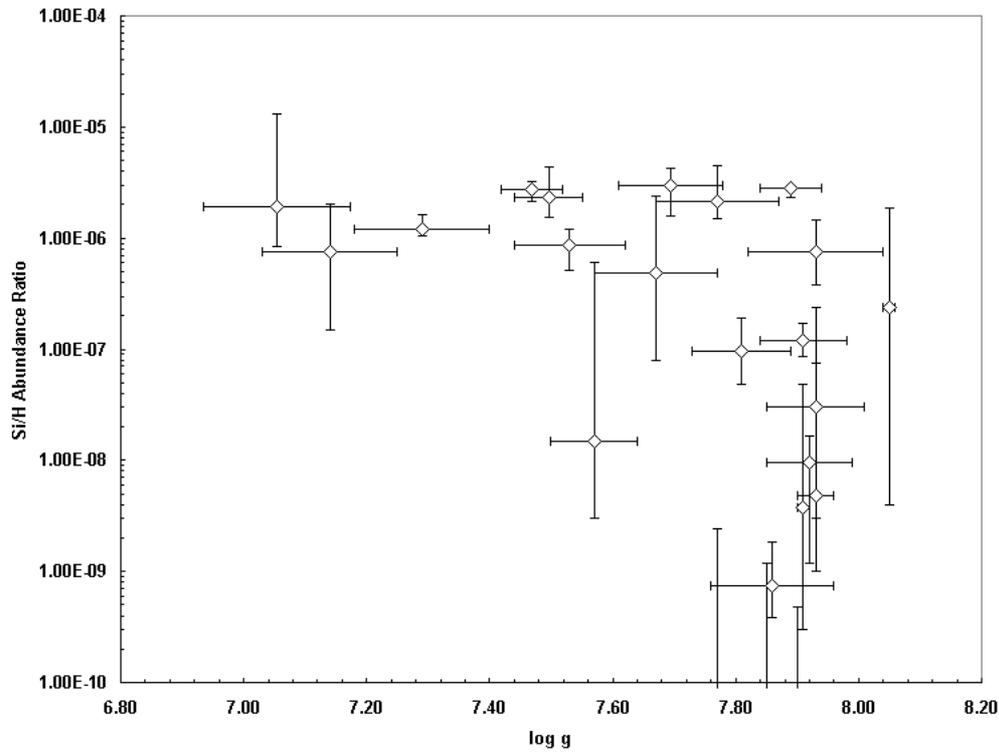

**Figure 15.** Measured abundances of Silicon (by number with respect to hydrogen) as a function of log $g$.

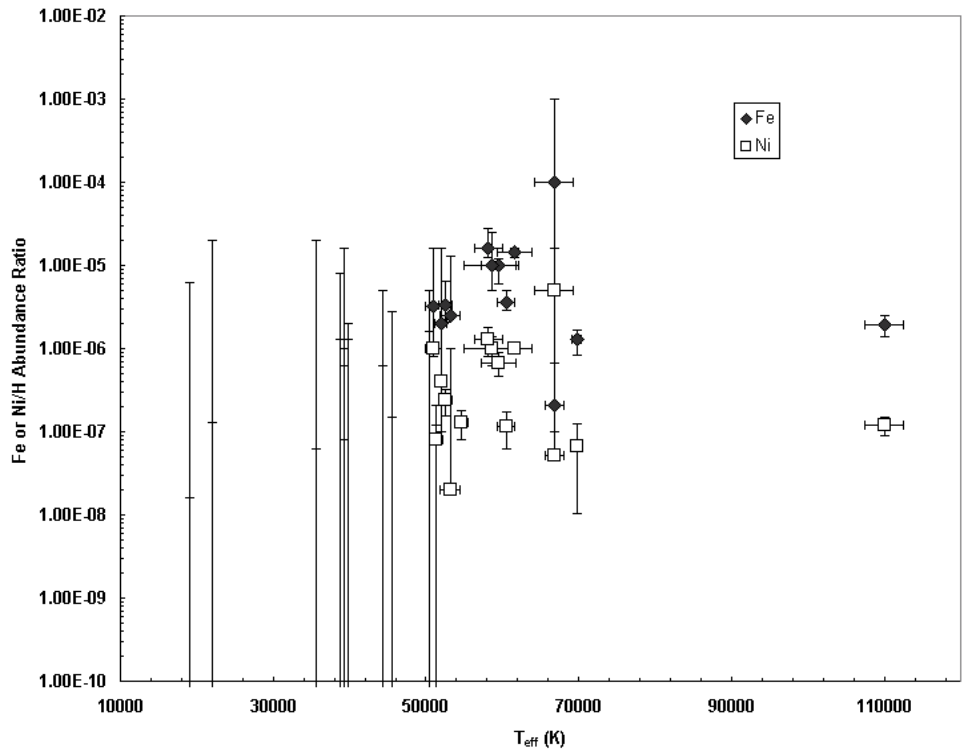

**Figure 16.**  Measured abundances (by number with respect to hydrogen) of Iron (filled diamonds) and Nickel (open squares) as a function of $T_{eff}$.

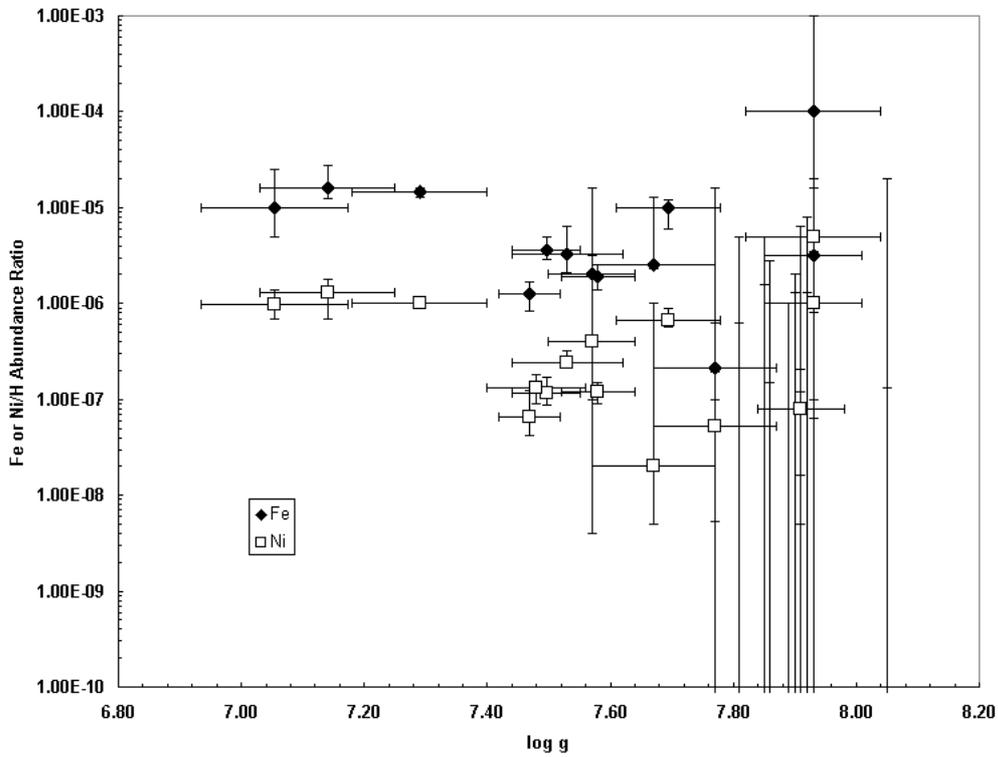

**Figure 17.**  Measured abundances (by number with respect to hydrogen) of Iron (filled diamonds) and Nickel (open squares) as a function of $log$ g.

*5.2.1 Carbon*

Abundances of carbon have been derived from measurements of CIII and CIV lines separately for most stars and these are displayed together in Figs. 8 and 9 to allow direct comparison. In some observations, with restricted wavelength coverage, only CIV was available. For individual stars, the two different measurements are largely in agreement, within the observational uncertainties. However, in the hottest group, there seems to be a systematic difference with the CIV measurements yielding larger abundances than for CIII. This might reflect a problem with the ionization balance, arising from the models themselves or an inappropriate choice of temperature. Alternatively, the respective lines may be sampling different atmospheric depths and, therefore, revealing the underlying carbon distribution.

As expected from our existing general understanding of how the composition evolves as white dwarfs cool, carbon is present in all the hottest stars and disappears from the atmosphere of many at temperatures below ~50000K. However, significantly, 3 of the sub-50000K stars (REJ1032+532, REJ1614-085, GD659) do have detectable abundances of carbon. Apart from REJ2156-546 the levels are very similar to those found in the higher temperature objects. In REJ2156-546, the carbon is only just detectable and the abundance 1 to 2 orders of magnitude below the typical range. At 35700K, GD659 is the lowest temperature object in the sample in which carbon is detected, but only EG102 and Wolf1346, at ~20000K, are cooler. Furthermore, the upper limits on the abundances exceed the values measured for other stars.

It has been clear for some time that the predictions of radiative levitation calculations, such as those of Chayer, Fontaine & Wesemael (1995a) and Chayer et al. (1994, 1995b) that are not included self-consistently in a model atmosphere calculation are of limited use. Typically, the detailed abundance predictions for an individual star of known temperature and gravity do not match the observations (e.g. Chayer et al. 1995; Barstow et al. 1997). Even so, across the white dwarf population the general patterns of element detections are reproduced. More recent, self-consistent treatments, such as those developed by Dreizler & Wolff (1998) and Schuh et al. (2002) have only been applied to EUV spectra and nothing has yet been published that we can compare with the UV data discussed here. Hence, the work of Chayer et al. (1994, 1995a, 1995b) still represents the best reference point, for the time being.

If we compare the data of Fig. 8 with the predicted run of abundance with temperature in Chayer et al. (1995b), it is evident that we might indeed expect to see measurable quantities of carbon down to $T_{eff}$~30000K. In this context, those stars that show no evidence of carbon in the 30000K to 50000K temperature range are more problematic than those that do. Above 50000K, we see no dependence of the C abundance with temperature, which is in agreement with expectations. In contrast, the expected dependence on gravity is not observed, but the uncertainties in the measurements are the same order of magnitude as the predicted abundance range.

*5.2.2 Nitrogen*

The nitrogen abundances (Figs. 10 & 11) follow a similar pattern to carbon. However, the abundances measured for 3 stars with $T_{eff}$ below 50000K (REJ1032+532, REJ1614-085 and GD659) are higher than any other measurements. Furthermore, the presence of flux from a near pure-H envelope in the EUV in REJ1032+532 and GD659 shows that the N present cannot extend to any great depth (down to about $3 \times 10^{-16} M_\odot$, Holberg et al. 1999). While the presence of significant abundances of N would be predicted by radiative levitation calculations, the high degree of stratification is not. Again, there appears to be a division of the cooler stars into two groups, with and without N in their atmospheres. Interestingly, there is a hint of a divergence between 50000K and 55000K where 3 objects show a clear trend of decreasing abundance with temperature, while two others have abundances similar to the higher temperature stars.

*5.2.3 Oxygen*

The abundances of oxygen (Figs. 12 & 13) are based on detection of the OV line. Hence, at the lower temperatures, where this feature is difficult to excite, the upper limits on the oxygen abundance are not very informative. Of the 3 stars showing C and N in this range only REJ1032+532 has a detection. However, the wavelength range of the REJ1614-085 observation did not include OV. It is important to note that there are no detections of OIV lines in those spectra where OV is not seen, although we do not quote formal abundance limits for these. No trends of abundance with temperature or gravity are noted. There is a general tendency to have more non-detections for higher gravity stars (also for C and N), but gravity increases as white dwarfs evolve to lower temperature and this may just reflect that trend. The radiative levitation calculations predict that oxygen should disappear from the white dwarf atmospheres at higher temperatures than either C or N, which is consistent with our results.

*5.2.4 Silicon*

The variation of abundance with temperature (Fig. 14) is very striking for silicon. The highest values are observed for the hottest stars, but there is then a strong decline in abundance from ~55000K down to ~40000K, with the exception of GD394, which contains an abundance similar to that of the hottest stars. However, we should probably discount GD394 from this discussion because of its known peculiarities (Dupuis et al. 2000). There is a hint of a rise in the Si abundance at temperatures below 40000K. However, the error bars of the Si abundance measurements are very large for these cooler objects. Nevertheless, it is interesting to note that Si is the only detected element, among those included in this study, in the two coolest stars in the sample.

In their calculations, *Chayer* et al. (1995b) do predict a minimum in the abundance of Si, but at ~70000K, some 30000K hotter than is observed here, if indeed we have observed a true minimum. Earlier discussions of the detection of Si in Wolf1346 and EG102 (Holberg et al. 1996, 1997) also include reference to the presence of Al and Mg. While the presence of Si can be readily accounted for by the radiative levitation calculations, Al and Mg should have disappeared from the photospheres of these two stars, indicating that there is ongoing accretion from some source. Such accretion may also be contributing to the photospheric Si.

*5.2.5  Iron and Nickel*

We have displayed the measured abundances of iron and nickel together in Figs. 16 and 17 to allow easy comparison of measurements for the only two iron group elements detected in the stars in this sample. While there have been hints of the presence of other Fe-group elements, such as Cr (Holberg et al. 1994), definite detections have yet to be reported. It is clear from both Figs. 16 and 17 that the abundances of Fe and Ni track each other very closely. Fig. 18 shows the ratio of the Fe:Ni abundances as a function of $T_{eff}$ for all those stars where both elements are detected. As the errors on the individual abundance measurements are not symmetrical, about the abundance value, we estimate the uncertainties in the ratio calculation by considering the possible extremes of the Fe/Ni ratio (i.e. Fe value+3σ/Ni value-3σ, Fe value-3σ/Ni value+3σ). Within the measurement uncertainties, the Fe/Ni ratio is more or less constant at a factor close to the cosmic value (solid line, ~20), whereas Chayer et al. (1994) predict a ratio close to unity.

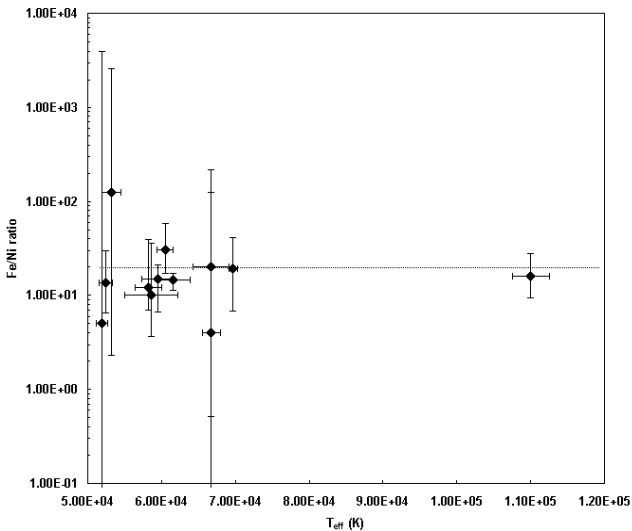

**Figure 18.** The ratio of observed Fe and Ni abundances calculated for those stars where both elements are detected. The error bars were computed as described in the text and the horizontal line marks the solar ratio, which has a value ~20.

The values of the Fe and Ni abundances show a strong bimodal dependence upon temperature. Below ~50,000K, neither of these elements is detected in any star in our sample. In contrast, all stars in the upper temperature range show both elements, with a fairly small scatter in abundances encompassed largely by the measurement uncertainties. Fig. 17 shows a trend for decreasing Fe and Ni abundance with increasing gravity. Variations in log *g* between individual stars could largely account for the scatter observed in the Fig. 18 abundance vs. temperature plot. We need to be a little cautious in interpreting the non-detection of Fe and Ni below 50000K, since FeIII has been detected in GD394 by *FUSE*, at shorter wavelengths than covered by the spectra discussed in this paper (Chayer et al. 2000). Our observations could be explained by the absence of transitions of lower ionization stages in the *HST* and *IUE* wavebands rather than a lack of Fe or Ni in the white dwarf atmospheres. However, as discussed earlier, GD394 has some peculiarities and this argument cannot explain the non-detections of Fe and Ni in $T_{eff}$~50000K stars like HZ43.

**5.3  STIS observations of stars with $T_{eff}$>50000K**

The STIS instrument on *HST* has provided the best signal-to-noise data utilized in this study. As a result, for those stars where STIS spectra are available, the abundance measurements uncertainties are much smaller than for the *IUE* observations. Consequently, it is a useful exercise to treat these stars as a high quality subsample in which we might see systematic evolutionary trends. We restrict this analysis to those stars where all the elements considered here have been detected. In Fig. 19, we show the abundances measured for each element for a total of 7 stars. The results are not very conclusive as, even with these most accurate measurements, the error bars overlap substantially. It would be reasonable to say that, within the measurement uncertainties, the observed abundances are the same in each star. Therefore, we are unable to detect a trend related to white dwarf cooling, as has been reported by Vennes et al. (2001) for Feige 24 and G191-B2B. Nevertheless, there are some weak general features to note. The measured abundances in WD2218+706 are the highest in the sample for most elements, whereas those for REJ1738+665 are the lowest. It is most likely that we are seeing a gravity effect, since WD2218+706 as by far the lowest surface gravity in the sample and REJ1738+665 has the highest surface gravity in this group.

**6  DISCUSSION**

In section 5 we have presented a complex analysis of the measured abundance of heavy elements for 25 hot DA white dwarfs observed at high spectral resolution in the far-UV. This sample represents the most complete survey of the evolution of the composition as white dwarfs cool yet published. In particular, we have examined how the abundances of key elements vary with stellar temperature and gravity, compared these results to the predictions of radiative levitation calculations and studied particular important subgroups of stars.

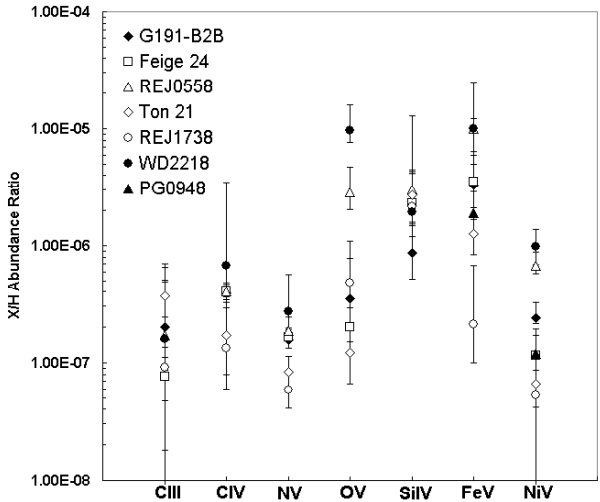

**Figure 19.** Comparison of measured abundances for all elements included in this study for all stars hotter than 50000K observed by STIS.

A key element of this study has been the careful objective approach taken to measure abundances with all available lines (or most for Fe and Ni) of a particular species treated simultaneously, using a $\chi^2$ goodness of fit, as described in section 4. This provides a firm statistical basis for determining the measurement uncertainties on which much of the analysis (section 5) and the discussion in this section is based. Importantly, it allows us to treat correctly the non-uniformity of the signal-to-noise in the observations, particularly when comparing *IUE* and *HST* data. The sizes of the uncertainties in the abundance measurements place clear limits on their interpretation and sometimes make it difficult to draw substantial and interesting conclusions. Nevertheless, some important key results have emerged. We reiterate each of these in turn below and then discuss them further in the subsequent subsections.

1. The presence or absence of heavy elements in the hot DA white dwarfs largely reflects what would be expected if radiative levitation is the supporting mechanism.
2. The measured abundances do not match the predicted values very well.
3. While all stars hotter than ~50000K contain heavy elements, as expected, there is an unexplained dichotomy at lower temperatures with some stars having apparently pure H envelopes and others detectable quantities of heavy elements.
4. Where heavy elements are present in stars with $T_{eff}$ below 50000K, they are often stratified, lying in the outermost layers of the envelope.
5. When detected, the Fe and Ni abundances maintain an approximately constant ratio, close to the cosmic value ~20.
6. For most of the hottest white dwarfs, the spread in element abundances is quite narrow and similar to the abundances measured in G191-B2B.
7. A few strong temperature/evolutionary effects are seen in the abundance measurements. These are: Decreasing Si abundance with temperature. N abundance split into two groups at lower temperature (see also 3). Sharp decline in Fe and Ni abundance to zero, below ~50000K.
8. For the hottest white dwarfs observed by STIS, the strongest determinant of abundance appears to be gravity.

*6.1.1 Radiative levitation and element abundances*
The role of radiative levitation as a mechanism for supporting heavy elements in a white dwarf atmosphere against the downward force of gravity is well established and we have discussed the most recent key papers by Chayer et al. (1994, 1995a, 1995b) earlier in this one. The general pattern of EUV opacity with stellar temperature and gravity, mapped out by photometric observations, has provided strong supporting evidence (e.g. Barstow et al. 1993; Marsh et al. 1997; Vennes et al. 1996). Conversely, the predicted abundances have usually been in conflict with measured values (e.g. Holberg et al., 1993,1994; Chayer et al. 1994, 1995a, 1995b). The work reported here underlines this very strongly.

Chayer et al. have acknowledged that there are deficiencies in their theoretical framework. More recently Dreizler & Wolff (1999) and Schuh et al. (2002) have made improvements in the approach by incorporating the radiative levitation calculations self-consistently into their atmosphere code. In particular, Schuh et al. have been very successful in matching the observed *EUVE* spectra of a sample of stars with synthetic spectra, using just $T_{eff}$ and log $g$ as the free parameters, from which the element abundances are determined within the models. From this work they derive a metallicity parameter, which shows a general decrease with decreasing temperature. However, there remain some potential problems and limitations. First, the values of $T_{eff}$ and log $g$ determined from their fits are not consistent with those independently measured from the Balmer lines. Secondly, the metallicity of stars that are quite different in appearance in the UV (e.g. REJ2156-546 and REJ1032+532) is found to be the similar from the EUV analyses. Thirdly, these new self-consistent models have yet to be tested against the high spectral resolution far-UV data, which contain much more detail than the EUV spectra.

*6.1.2 Additional abundance altering mechanisms?*
The key result that must be explained by any new theoretical work is our observation of the different heavy element abundance levels in otherwise similar stars below ~50000K. This is in contrast to the fairly uniform abundances seen at higher temperatures. This might in part be explained by uncertainties in the various measurements, but there are striking examples of pairs of stars (e.g. RE2156-546/REJ1032+532; REJ1614-084/GD153) where the difference in abundance is well outside these possible errors. It seems likely that some additional effect (or effects), over and above radiative levitation, is (are) at play here. There are several possibilities that must be considered, which might operate individually or in some combination.

In some stars, material might be being accreted from a companion, from the ISM or from previously ejected stellar material. The stratification effects seen in REJ1032+532, REJ1614-085 and GD659 would lend support to this idea. However, only REJ1032+532 has been noted as a possible member of a binary system (Holberg et al. 1999). Furthermore, EUV spectroscopy shows that both REJ1032+532 and GD659 lie in low-density regions of the ISM, indicating the accretion from the ISM will not be significant. Evidence for circumstellar material has been seen in some of the hottest white dwarfs and such features are seen in REJ1614-085 (Bannister et al. 2003). There is also a weak detection in GD659, but nothing is seen in REJ1032+532. While, we cannot identify a clear single explanation that might work for all three stars, there are possible mechanisms for each.

*6.1.3 Why do some white dwarfs have pure H envelopes?*
Coupled with the above discussion is the need to explain the existence of a number of stars that do not contain heavy elements in their photosphere: HZ43, GD153 and PG1057+719. The radiative levitation calculations and the fact that other stars with similar temperature and gravity do have such photospheric material is a problem. Some of the mechanisms discussed in 6.1.2 might explain the differences between the pure H and metal-containing groups, but it is also important to examine the possibility that the pure H stars instead may be peculiar in some way. All the radiative levitation and diffusion calculations assume the presence of a reservoir of heavy elements and uniformly distributed abundances are usually the starting point. However, there is no a priori proof that such a reservoir must be present. For most white dwarfs, nuclear fusion products are limited to nothing more massive than C and O. Hence, all heavier elements (and probably N) must be derived from the composition of the progenitor. Since all the white dwarfs studied here are local disk objects it seems very unlikely that any of the progenitors could have been metal poor. Therefore, we need to look for a mechanism that depletes the heavy element reservoir. Mass-loss through a wind might do this. However, for radiatively driven winds, mass loss should be lower in the pure-H stars than those containing heavy elements, because the wind is driven by metal lines

*6.1.4 The scope of the DA sample*
The sample of 25 white dwarfs assembled here is the largest for which high-resolution far-UV spectra have been examined in such detail. Nevertheless, it has a number of limitations. First, it is an assembly of small groups of observations from a number of separate programs. These have used different instruments, yielding a range of signal-to-noise, and some have incomplete spectral coverage. Secondly, the target selection criteria have been very non-uniform. Hence, the sample is not necessarily representative of the DA population as a whole. For example, many of the *IUE* observations were obtained as part of a project to identify heavy elements in stars already known to have strong EUV opacity and those appearing as pure H atmosphere in the EUV were ignored. Interestingly, the STIS observations of those key objects in our discussion, REJ2156-546, REJ1032+532 and GD659, were part of a program to study the local ISM. Without these spectra, the abundance dichotomy we observe below 50000K would be much less clear, relying just on REJ1614-085, which might easily have been dismissed as a single peculiarity. Of course, with only small numbers and rather biased selection, we do not know whether or not REJ1032+532, REJ1614-085 and GD659 are typical of DAs in this temperature range or if true pure H envelopes like GD153 and HZ43 are "normal".

*6.1.5 The scope of the far-UV observations*
These far-UV spectra represent the most detailed information available in any spectral range for these stars. However, they do not necessarily tell the whole story. For example, for the coolest white dwarfs the Nv lines will not be excited and there are no other nitrogen transitions in the 1150-1900Å range. Hence, nitrogen could be present but not detected for this reason. Observations at shorter wavelengths, in the *FUSE* band, provide access to some lower ionization stages of N and other elements. FeIII has been detected in the *FUSE* spectrum of GD394 (Chayer et al. 2000) but only higher ionization stages of Fe are present in the longer wavelength spectra.

We have also invoked our knowledge of the EUV spectra of some stars in interpreting the observations presented here. Without this data, we would have a rather different view of some of the stars in our study. It is clear then, that a multi-wavelength approach, of which far-UV spectra are an essential part, is the only way of completely understanding the complexities of individual stars.

We have drawn data from two major UV telescopes, *IUE* and *HST*. However, it is of variable quality. The spectral resolution of the *IUE* is approximately 3-4 times poorer than that of the *HST* M gratings of the GHRS and STIS instruments. The circumstellar material detected in some objects like G191-B2B (see e.g. Bannister et al. 2003) would not have been separated from the photospheric lines at the resolution of *IUE*. Had they been observed by *IUE*, it is likely that we would have reported higher abundances than tabulated in this paper. Only in REJ0457-281 have circumstellar absorption components been detected and resolved by *IUE*. Therefore, we should be concerned as to whether or not undetected components may be present in other stars observed only by *IUE*, such as REJ2214-492 and REJ0623-371. In fact, we cannot necessarily assume that all such components have been resolved by the STIS E140M grating. For example, the circumstellar material in G191-B2B was barely resolved in the C IV lines. Problems in fitting model atmosphere line profiles to the observations of GD246 and PG0948+534 could indicate that there is circumstellar material present, but no direct evidence for any multiple absorption components is found by Bannister et al. (2003). Therefore, if present, circumstellar material must lie at a velocity so close to that of the photosphere that it is beyond present spectral resolution capabilities.

*6.1.6  Scope of the models*
All the results presented here depend on the stellar model atmosphere calculations and the detailed physics and atomic data included therein. However, all model calculations have some limitations, none are perfect. Better treatments of particular parts of the problem are being developed all the time. Indeed, even the TLUSTY code used here has undergone further recent development. It would be hugely time-consuming and probably unnecessary to repeat our analysis for each new generation of the codes. What is most important is that the analyses have been carried out using models calculated with the same version of the program. Therefore, the internal comparison of abundances on an object-by-object basis is valid. Furthermore, it is clear from this work that, even with the best quality spectra, the statistical uncertainties in the abundance measurements are significant. At the moment, they are probably the dominant limitation of this work, rather than the model systematics.

*6.1.7  Other limitations of the analysis*
While we have tried to improve the objectivity of the stellar abundance measurements, with the statistical technique employed. There remain several limitations that cannot be ignored. For example, we have only considered the statistical errors in the analysis and failed to consider any systematic effects. For example, it is clear from work published in several papers that the actual uncertainties in measurements of $T_{eff}$ and log $g$ are larger than the statistical values reported (e.g. Napiwotzki et al. 1999; Barstow et al. 2001). Moreover, we have made necessary simplifying assumptions about the general composition of the stars when carrying out the Balmer line analysis. Ultimately, the study of any individual star needs to be completely self-consistent, where the model of the appropriate abundance is used for the Balmer line work. Furthermore, in all our analyses, we have just dealt with a single species at any one time. However, it would be much better to deal simultaneously with all ionization stages of an atom and even with all atoms together. The consequence of these limitations is that the true abundance uncertainties are large than the values that we quote here, although the fact that we quote 3σ values, rather than the standard 1σ errors, will mitigate this to some extent. A completely self-consistent abundance, temperature and gravity analysis is the ideal, but beyond the computational scope of current resources.

**7  CONCLUSION**

We have reported a series of systematic abundance measurements for a group of hot DA white dwarfs in the temperature range 20000-110,000K, based on far-UV spectroscopy with *HST*, *IUE* and *FUSE*. Using our latest heavy element blanketed non-LTE stellar atmosphere calculations we have addressed the heavy element abundance patterns making completely objective measurements of abundance values and upper limits using a $\chi^2$ fitting technique.

We find that the presence or absence of heavy elements in the hot DA white dwarfs largely reflects what would be expected if radiative levitation is the supporting mechanism, although the measured abundances do not match the predicted values very well, as reported by other authors in the past. Almost all stars hotter than ~50000K contain heavy elements. For most of these the spread in element abundances is quite narrow and similar to the abundances measured in G191-B2B. However, there is an unexplained dichotomy at lower temperatures with some stars having apparently pure H envelopes and others detectable quantities of heavy elements. The heavy elements present in these cooler stars are often stratified, lying in the outermost layers of the envelope. The inability to fit the C, N, O and Si resonance lines of PG0948+532 and the C IV lines of GD246 may also be evidence for stratification. A possible alternative explanation for all these stars is that there is circumstellar material present but at a velocity close to that of the photosphere, in contrast to those stars where circumstellar components are resolved (Bannister et al. 2003).

Only a few strong temperature/evolutionary effects are seen in the abundance measurements. There is a decreasing Si abundance with temperature, the N abundance pattern splits into two groups at lower temperature and there is a sharp decline in Fe and Ni abundance to zero, below ~50000K. When detected, the Fe and Ni abundances maintain an approximately constant ratio, close to the cosmic value ~20. For the hottest white dwarfs observed by STIS, the strongest determinant of abundance appears to be gravity.

Although we have obtained several interesting results from this work, it is clear that the ultimate scope of our conclusions is limited by the quality of the data, the number of stars observed and the wavelength ranges considered. Further progress can only be made by improving the data and expanding the sample. We need to eliminate some of the biases in the stars chosen for far-UV observations. Most observations have concentrated on those objects where there was already evidence for photospheric heavy element opacity and these stars have received good exposures to search for weak iron group lines. This is not the case for other stars, where better exposures are required to ensure that heavy element measurements have usefully small uncertainties or that the upper limits on abundances are meaningful. The number of stars with temperatures below 50000K needs to be expanded in an unbiased way, to decide whether either of the pure H or metal-containing groups we have identified are typical for the DA population or if there really are two distinct lines.

All this work needs to be carried out with a consistent multi-wavelength approach. *FUSE* spectra are probably better suited to measuring abundances of lower ionization stages of some important elements such as N or Fe. Fortunately, many of the targets in our sample have already been observed by *FUSE*. The presence of as yet undetected companion stars may be responsible for some of the abundance patterns we see. However, we do not have a detailed picture of the incidence of binarity in these stars. The signatures from cool M and L dwarfs can easily be masked by the hot white dwarfs at visible wavelengths. Therefore, a program of infra-red photometry and spectroscopy is required to check for the presence of companion stars. We have initiated such a programme, for the northern stars, on UKIRT (Burleigh et al. 2003). Much of our analysis and interpretation has been informed by knowledge of the *EUVE* spectra of some of our sample stars. However, these spectra are limited in resolution and number. Higher resolution and higher signal-to-noise access to the EUV will be an essential element of any future studies.

Finally, it is clear that much depends on the continuing availability of high signal-to-noise far-UV spectra. These have been the key to the work presented here. In particular the high resolution is essential to ensure that ISM, circumstellar and photospheric components are not confused. There will always be a practical limit to what is available, but the contrast between the *IUE* and STIS capabilities demonstrates that this should be at least at the resolving power (~40000) of the E140M grating and preferably the E140H (~100000). It is worrying that no such future UV capability is guaranteed.

**ACKNOWLEDGEMENTS**

The work reported in this paper was based on observations made with the *IUE, HST* and *FUSE* observatories. Extensive use was made of the MAST archive and the NEWSIPS processing of the *IUE* data. We would like to thank Ms Cherie Miskey for her assistance in processing the STIS data. MAB, MRB and NPB were supported by PPARC, UK. JBH wishes to acknowledge support provided by NASA through grant AR-9202, from the Space Telescope Science Institute, which is operated by AURA incorporated under NASA contract NAS 5-26555. FCB acknowledges the use of the computational facilities at the Laboratory for Astronomy and Solar Physics at NASA/GSFC.